\documentclass[lnbip]{svmultln}

\usepackage{graphicx}             	    
\usepackage{amsmath,amssymb}
\usepackage{multirow,multicol} 
\usepackage[bmargin=1.2in]{geometry}  
\usepackage{xcolor}
\usepackage{smartdiagram}
\usepackage{float}
\usepackage{caption}
\usepackage{subcaption}

\newcommand{\figref}[1]{\mbox{Figure \ref{#1}}}
\newcommand{\tabref}[1]{\mbox{Table \ref{#1}}}
\newcommand{\secref}[1]{\mbox{Section \ref{#1}}}
\graphicspath{{figs/}}

\begin{document}

\title{Revealing the research landscape of Master's degrees \\via bibliometric analyses}
\titlerunning{Revealing the research landscape of Master's degrees}

\author{Nathalia Chaparro\inst{1} \and Sergio Rojas--Galeano\inst{1}}
\authorrunning{Chaparro and Rojas--Galeano} 
\institute{
              $^1$Universidad Distrital Francisco José de Caldas, Bogotá, Colombia. \\
              \email{enathaliach@gmail.com}\\ 
              \email{srojas@udistrital.edu.co}\\ 
}

\maketitle    

\begin{abstract} 
\textcolor{black}{The evolution of a Master's programme, like many other human institutions, can be viewed as a self-organising system whose underlying structures and dynamics arise primarily from the interaction of its faculty and students. Identifying these hidden properties may not be a trivial task, due to the complex behaviour implicit in such evolution. Nonetheless, we argue that the programme's body of research production (represented mainly by dissertations) can serve this purpose. Bibliometric analyses of such data can reveal insights about production growth, collaborative networks, and visual mapping of established, niche, and emerging research topics, among other facets. Thus, we propose a bibliometric workflow aimed at discovering the production dynamics, as well as the conceptual, social and intellectual structures developed by the Master's degree, in the interest of guiding decision-makers to better assess the strengths of the programme and to prioritise strategic goals. In addition, we report two case studies to illustrate the realisation of the proposed workflow. We conclude with considerations on the possible application of the approach to other academic research units.
}


\end{abstract}
\section{Introduction}

\textcolor{black}{In most countries, master's degrees are academic programmes in which students are trained in specialised knowledge and then must complete a dissertation on a given research topic under the guidance of a faculty supervisor. Perhaps it is the fact that dissertations are carried out as a teamwork and knowledge-oriented activity, within a decentralised system, what conveys this type of academic program with the typical features of a complex system \cite{jacobson2019education, woolcott2021partnered}.
}

\textcolor{black}{This assumption can naturally be transferred to a scientific community. As with other social institutions that self-organise to face the uncertainties found in their environments \cite{arevalo2015theoretical}, the research activity of a master's program exhibits emergence of conceptual and collaborative structures along with dynamics of continuous change and innovation that renders a research landscape difficult to identify directly. Nonetheless, we argue that the body of its research output, mainly in the form of dissertations, are the building blocks of its scientific development, one that can be examined through the lens of bibliometric methods in an attempt to understand how such landscape has evolved. In this sense, bibliometrics can be seen as a particular type of data-mining \cite{van2002discovery}, here tailored to discover patterns that help to explain its complex academic behaviour.
}

\textcolor{black}{Bibliometric techniques provide useful information on the production and consumption of academic production in a framework of impartial, systematic and reproducible analysis for a given bibliographic corpus. The source, context and extent of the corpus will define the purpose and unit of analysis of the bibliometric study. Therefore, it is possible to perform this type of analysis to study the behaviour of a variety of academic units, including journals \cite{Donthu2020, Verma2015, Ramasamy2017, Das2013, Suarez2019, Lopez2018}, individual authors \cite{Ain2019, Nosek2010, Ben2010, Mingers2009}, scientific disciplines \cite{garousi2016citations, Merigo2019, Parlina2020, Paiva2020}, emerging topics \cite{Chahrour2020, Torres2020, Saed2017}, universities \cite{Cancino2017, Pradhan2018, Tarrio2017}, university programs or departments \cite{Nishat2019, Mondal2018, Eckel2009, Kelly2015} and even nation-wide assessments of specific thematic disciplines \cite{Hsieh2013, Yu2018, Jalal2019, rosselli2021bibliometric}.
}

\textcolor{black}{
When applied to a master's programme, such analysis may lead to a critical appraisal of academic production in terms of its bibliometric performance, as well as of the development of conceptual, intellectual and social structures of its associated research activity. Insights into production growth, faculty and group engagement, dominant and emerging topics of interest, collaboration patterns, and intellectual structures can convey useful information to decision makers such as the programme leaders, internal or external evaluators, faculty members and enrolled or future students. 
}

\textcolor{black}{
This study proposes a bibliometric workflow  to help reveal the research landscape of a Master of Science degree, using a multifaceted analysis based on its structural and dynamical properties. The application of the workflow is illustrated in two case studies of master's degree programs in engineering. The paper begins with a brief literature review of related works (Section 2), followed by an overview of the workflow (Section 3) and a detailed description of the stages involved in it (Section 4). We then report the results of the case studies (Section 5). The document concludes by discussing some ideas for future work.
}

\section{Related work}

Numerous studies related to the bibliometric analysis of different academic units have been published. For example, \cite{Tarrio2017} reports the analysis of the research output of a group of three universities in Spain, including descriptive and impact metrics 
to identify the elite of most productive authors on each university. Another work that explores the production, impact and collaboration of researchers in Information Sciences in Latin America and the Caribbean was carried out using bibliometric techniques \cite{Sanchez2017}. A more recent study proposes an approach to exploring the major themes of a text collection to obtain thematic mappings, with application to Big Data \cite{Parlina2020}. 

In contrast, our work focuses on the analysis of graduate school production. In this regard, \cite{Paiva2020} presents a study with quantitative indicators and a conceptual map obtained from dissertations and theses on chronic diseases in Brazil. Similar works have been reported evaluating the impact of citations, research topics and preferred journals for the publication of results for the Department of Library Sciences of the University of Calcutta \cite{Nishat2019}, or for the doctoral thesis in Mathematics and Political Science at Burdwan University \cite{Mondal2018, Mondal2017}. A study concerning a scientometric analysis of doctoral theses on the subject of Roma people, has recently been published \cite{Salgado2020}. 

In the same vein, our work describes the application of several performance and scientific mapping techniques to a bibliographic dataset of dissertations (we are not introducing either any novel bibliometric technique), but differs in that instead of a quantitative vs qualitative approach, we outline a generic workflow for analysing structures and dynamics of knowledge, where a variety of techniques are explicitly aimed at discovering specific patterns describing the general picture of the research landscape. This approach is the main contribution of this paper and is explained in \secref{sec:M&M}.

The realisation of the workflow can be carried out using any bibliometric or scientific mapping software tool that supports the chosen techniques. Several tools have been applied in the reviewed literature: \textit{VOSviewer} \cite{Donthu2020, Merigo2019,Cancino2017,Saed2017,Qiu2014}, \textit{Bibexcel} \cite{Sanchez2017}, \textit{Taverna} \cite{Guler2016}, T-LAB \cite{Parlina2020}.
However, in this regard, we decided to use \textit{bibliometrix}, an open source R library for full bibliometric analysis and scientific mapping  \cite{Aria2017}. Since its recent introduction, this toolkit has been widely adopted by the community to perform a variety of bibliometric analyses in various disciplines (see  \cite{Jalal2019, Dervics2019, Nafade2018, Brito2018, Javid2019, Salgado2020, Campra2021, Aria2020, Fortuna2020, Brito2018, Javid2019, Suarez2019, Warin2020}, to name a few). 

Nonetheless, we note that our approach is tool-independent and therefore any other software option (or combination of software tools) can be used as long as they are compatible with the techniques involved. For a complete review of this type of tools, we refer the reader to \cite{Moral2020} and references within.

\section{Workflow description}\label{sec:M&M}

\subsection{Overview}
\begin{figure}[h!]
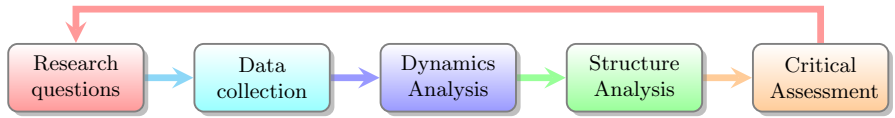

    \smartdiagramset{distance center/other bubbles=1.2cm}
    \begin{center}
    \scalebox{0.9}{
        \smartdiagram[flow diagram:horizontal]{
         Research\\questions,
         Data\\collection, 
         Dynamics\\Analysis, 
         Structure\\Analysis, 
         Critical\\Assessment
        }
    }
    \end{center}
    \caption{The proposed workflow.}
    \label{fig:method}
\end{figure}

\textcolor{black}{
The workflow consists of the stages depicted in \figref{fig:method}. The initial stage encompasses the definition of the questions that guide the analyses aimed at discovering the research landscape of the master's programme. We have identified four key questions that we consider relevant for this aim, although these may be tailored to the specific targets of any particular study. Those questions are described in \secref{sec:questions}.
}

\textcolor{black}{
The next stage focuses on data collection. Here, once an observation window is defined, the metadata of the dissertations submitted within it is collected from institutional repositories or abstract and citation databases (and, optionally, papers derived from them). A detailed schema of the metadata to be collected is described in \secref{sec:dataset}.
}

\textcolor{black}{
From the collected bibliographic corpus, the following stages correspond to the actual analyses, which include performance bibliometrics and scientific mapping in an attempt to delve into the evolution of the dynamics and structural facets of the master's programme. The insights obtained from each of these facets would provide an enriched evaluation of its production behaviour during the observation window. The bibliometric techniques that would be used to feed these facets are described in \secref{sec:bibliometrics}.
}

\textcolor{black}{
In the last stage, a final critical assessment is made based on the findings of the previous bibliometric analyses; the aim 
would be to interpret the ideas about the dynamic and structural patterns discovered from the master's program, in a unifying reflective perspective that can provide useful information for strategic planning and decision-making. Furthermore, in view of the continuous evolution of the programme as a self-organising entity, the workflow can be applied routinely to account for such changes (a loop indicated by the red arrow in the figure).
}

\subsection{Research questions}\label{sec:questions}

\textcolor{black}{
We propose to focus the analyses on the two key properties that facilitate the appearance of complex behaviours in most human organisations: dynamics and structure \cite{alaa2009derivation}. Therefore,  we defined four central questions to address such aspects (RQ1 to RQ4, see \tabref{tab:questions}). RQ1 deals with the dynamics of the programme's production, from the point of view of performance: indicators of growth, impact and activity of research output. RQ2 and RQ3, in turn, are related to the emerging structures that support the research activity of the program. They were divided in two, RQ2 centred around the development of knowledge structures, while RQ3 focused on the emerging interaction between the actors that influence the production of research. Lastly, RQ4 is a synthetic question, the answer to which would be a reflection on the findings obtained in the other three questions, so as to provide an overall critical assessment and perspective of the research landscape obtained from the application of the workflow. 
}

\textcolor{black}{
Although we outlined these questions as a guidance intended to capture the broader picture of the emergent properties of the master's degree, we remark that they can be adjusted to other specific purposes (for example, comparing how the structures or dynamics have change with respect to an older analysis previously made).
}

\begin{table}[t]
    \footnotesize
    \centering
    \setlength{\tabcolsep}{.2cm}
    \def\arraystretch{1.2}
    \begin{tabular}{p{.6cm}|p{5cm}|p{8cm}}
    \hline
          \textbf{Id.} 
        & \textbf{Research question} 
        & \textbf{Motivations} \\\hline
 
         RQ1. 
         
         & How to characterise the scientific production dynamics of the Master's programme during the observation window?
         
         & To identify indicators of publication growth, citation impact and dynamics of the scientific production originated in the research projects carried on by groups during the time frame, in summary, the overall research performance. 
         \\\hline 
         
        RQ2. 
        
        & What are the distinctive features of the conceptual structures developed within the Master's programme during the observation window?
        
         & To discover frequently used index terms, dominant and emerging topics and thematic areas of research undertaken by groups and faculty associated with the Master's programme. 
         \\[-2ex] \hline 
         
         RQ3. 
         
         & What are the characteristics of the collaboration structures emanated within the Master's programme during the observation window?
         
         & To reveal the patterns of collaboration implicitly evolved within the Master's programme, considering social and intellectual networks of authors, groups, and common literature couplings.  
         \\[-2ex] \hline 
         
         RQ4. 
         
         & What are the critical factors the board of directors should prioritise in order to strengthen the performance of the Master's degree scientific landscape in the near future?

         & To assess the current state and outlook of established and emerging areas of research according to the strengths and weaknesses identified with the analysis conducted in the previous questions, so as to recommend  actions aimed at improvement of the scientific production structures and dynamics of the Master's programme.
         \\ \hline 

    \end{tabular}
    \caption{Research questions and motivations for the study design stage.}
    \label{tab:questions}
\end{table}

\subsection{Dataset collection}\label{sec:dataset}
\begin{figure}[t]
    \centering
    \includegraphics[scale=.4]{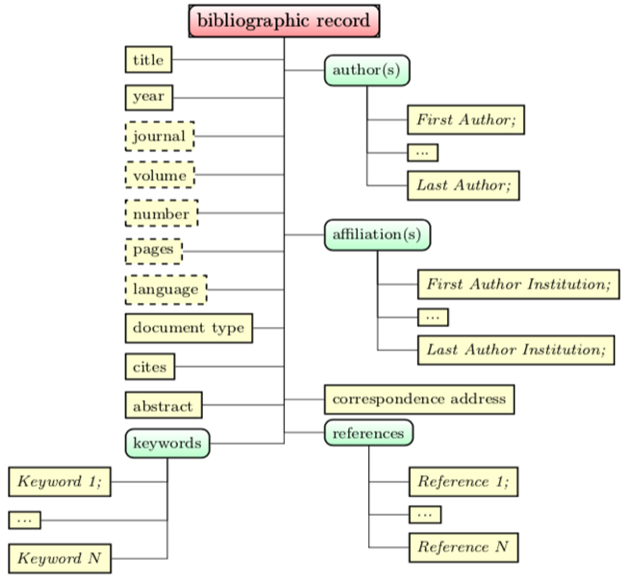}
    \caption{Scheme of the bibliographic record to assemble the corpus. Optional fields are denoted with dashed lines, single-value fields are depicted with sharp-corner rectangles, and compound fields are shown with rounded-corner rectangles. The latter correspond to lists of  semicolon-separated values.}
    \label{fig:metadata}
\end{figure}

\textcolor{black}{
To carry out the analyses described in the following section, first a data set must be assembled with bibliographic records of those dissertations defended during the observation window. To do this, we recommend organising the metadata corresponding to each record in the scheme shown in \figref{fig:metadata}. This scheme is designed according to the BIB format used by the BibTeX reference manager (see \cite{Fenn2006}). We believe that it is a convenient format because it is available as an export interface in most bibliographic databases such as Web of Science, Scopus, Google Scholar, institutional databases, and also in most reference management programs.
}

\textcolor{black}{
Although the fields in the BIB record were intended to primarily describe the metadata associated with papers, they can be reinterpreted to contain analogous information related to dissertations. For example, the field \textit{journal} can be associated with the research group or laboratory to which the student joined; the actual name of the student can be \textit{first author}; the list of other \textit{authors} may contain the names of supervisors and advisers, and similarly with the list of \textit{affiliations} (useful for external advisors). As we will see later (in \figref{fig:flow-metadata}), the data in each of these fields would be used in one or more of the bibliometric analyses that are described in the next section.
}

\textcolor{black}{
In addition to assembling bibliographic records into a .BIB file, we suggest doing some data cleaning, which involves removing typos, repeated or joined words, and symbols not recognised by the ANSI UTF-8 standard encoding. In the case of dissertations submitted in Spanish, we suggest removing accents and other punctuation marks in titles and names (such as \textit{á, é, í, ó, ú, ñ}, etc.), to improve the accuracy of the software tools used to perform the analysis. Moreover, since most bibliometric techniques use algorithms for natural language preprocessing, it is also recommended to configure stop-words and synonyms lists to filter non-informative or redundant terms, which can help to obtain more accurate results. Links to the .BIB files and lists that we used in the case studies reported in this study are provided in the Supplementary Material section.
}

Lastly, we remark that metadata of papers derived from dissertations could also be collected and organised as a complementary dataset using the same BIB record of \figref{fig:metadata}, and likewise, it could be used as input for an additional analysis of the properties described in the following section.

\begin{table}[bt]
    \scriptsize
    \centering
    \setlength{\tabcolsep}{.2cm}
    \def\arraystretch{1.2}
    \begin{tabular}{p{3.6cm}|p{6.5cm}|p{3.2cm}}
    \hline
        \textbf{Bibliometric technique} 
        & \textbf{Description} 
        & \textbf{Type}
        \\ \hline
        
    Production statistics 
         
         & Statistics of annual scientific production, average citations and other impact indicators. 
         
         & Descriptive analysis (\textbf{P})
         \\[-2ex] \hline 
         
    Production growth 
         
         & Plot of curves representing production counts arranged by year. 
         
         & Trend analysis (\textbf{P})
         \\[-2ex] \hline 
         
    Production distribution  
         
         & Frequency histogram of total dissertations per authors or groups.

         & Descriptive analysis (\textbf{P})
         \\[-2ex] \hline 
                  
    Citation count 
         
         & Plot of curves representing citation counts (total or averaged) arranged by year. 
         
         & Trend analysis (\textbf{P})
         \\[-2ex] \hline 
         
    Citation distribution  
    
         & Frequency histogram of citations for dissertations (total or yearly).
 
         & Descriptive analysis (\textbf{P})
         \\[-2ex] \hline 
         
    Author's timelines
         
         & A stack of 1D bubble diagrams representing dynamics and frequency of author's production (or also groups) over individual timelines.
         
         & Trend analysis (\textbf{P})
         \\[-2ex] \hline 

    Word trends
         
         & Plot of word usage trends over the years, obtained by title, abstract or keywords. 
         
         & Trend analysis (\textbf{P})
         \\[-2ex] \hline 
         
    Frequent words
         
         & Frequency histogram of words appearance, obtained by title, abstract or keywords. 
         
         & Descriptive analysis (\textbf{P})
         \\[-2ex] \hline 

    Word cloud
         
         & Cloud-shaped visual design of most frequent words, obtained by title, abstract, keywords.
         
         & Descriptive analysis (\textbf{P})
         \\[-2ex] \hline 

    Topic map
         
         & A plot where the proximity of words co-occurring in multiple documents is depicted in a 2D map as clusters defining topics or concepts.
         
         & Conceptual structure (\textbf{M})
         \\[-2ex] \hline 
         
    Word dendrogram
         
         & An alternative visual format to depict proximity of word co-occurrence, using a hierarchical tree displaying level-dependant partitions.
         
         & Conceptual structure (\textbf{M})
         \\[-2ex] \hline 

    Co-occurrence network
         
         & A network plot representing different features of words and relationships among them: co-occurrence, dominance and similarity clusters.  
         
         & Conceptual structure (\textbf{M})
         \\[-2ex] \hline 

    Thematic map
         
         & By clustering words according to centrality (importance in the field) and density (development in the field) a 2D map can be generated depicting  motor, emerging, declining and fundamental themes.  
         
         & Conceptual structure (\textbf{M})
         \\[-2ex] \hline 
         
    Collaboration network
         
         & Network of co-authorship patterns revealing collaboration links between authors, supervisors and groups.   
         
         & Social structure (\textbf{M})
         \\[-2ex] \hline 

    Authors coupling network
         
         & Network of authors connected if they share references cited in the entire oeuvres bibliography (their lists of supervised thesis).   
         
         & Social structure (\textbf{M})
         \\[-2ex] \hline 
                  
    Co-citation network
         
         & Networks of co-occurrence of citations, revealing structures of literature and authorship relevance.   
         
         & Intellectual structure (\textbf{M})
         \\[-2ex] \hline 

    Manuscript coupling network
         
         & Network of dissertations that are linked when they refer to shared works in their bibliographies.   
         
         & Intellectual structure (\textbf{M})
         \\[-2ex] \hline 

    Energy flow diagrams
         
         & Visual representation of energy exchange, i.e. the outflow and inflow of contributions, between bibliographic units (also known as Alluvial diagrams). 
         
         & Conceptual, intellectual, and social structure (\textbf{M})
         \\[-2ex] \hline 
         
    \end{tabular}
    \caption{A set of bibliometric techniques suggested to carry out the dynamics and structure analyses of the proposed workflow. The type of technique is associated to the bibliometric assessment they perform (\textbf{P}: \textit{Performance bibliometrics}; \textbf{M}: \textit{Science mapping bibliometrics}). }
    \label{tab:tools}
\end{table}

\subsection{Dynamics and structure analyses} \label{sec:bibliometrics}

\textcolor{black}{
According to \cite{noyons1999integrating}, the two main branches of bibliometric assessment are performance evaluation and scientific mapping. In line with that vision,  we designed the stages of dynamics and structure analyses to take advantage of the variety of techniques that are usually applied in each of these facets. By combining these two types of complementary analysis, we aim to build a more comprehensive assessment of the emerging research landscape of a master's programme. 
}

\textcolor{black}{
Having this in mind and taking into account the research questions defined above, we  chose a broad set of bibliometric techniques to apply in each analysis. On the one hand, we link the dynamics analysis with performance bibliometrics, where we consider growth, distribution and descriptive statistics of research production, author's timelines, as well as trends of terms, frequent words and word clouds. Additionally, citation counts and distributions were also included as a measure of visibility; in this respect, other bibliometric impact indicators, such as the  h-index and variants \cite{alonso2009h} were not included, since citation impact is not considered a typical outcome of a dissertation. The descriptions and categories of these techniques are summarised in \tabref{tab:tools}.
}

\textcolor{black}{
On the other hand, we associate the stage of structure analysis with science mapping bibliometrics, choosing techniques such as topic maps, word dendrograms, co-occurrence networks, thematic maps, collaboration and co-citation networks. These are also described in \tabref{tab:tools}, along with the type of  analysis they perform. A final technique was added, energy flow diagrams, which are useful for visualising aspects of both the dynamics and structure facets of the workflow.
}

\begin{figure}[t]
    \centering
    \includegraphics[scale=.41]{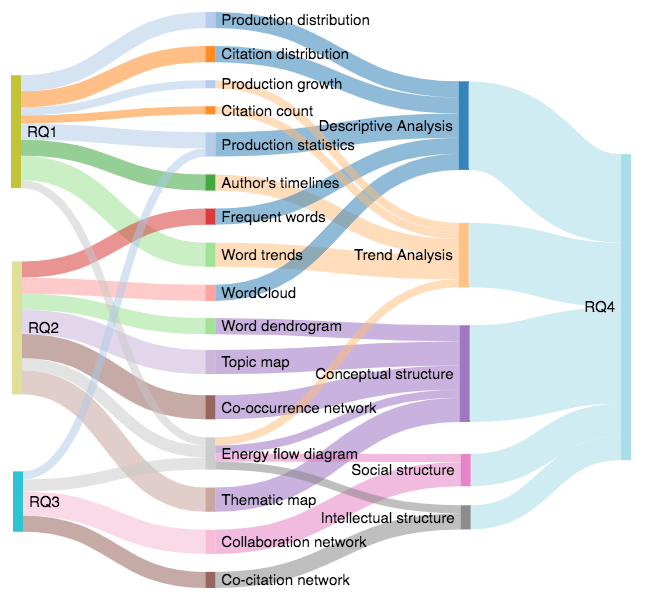}
    \caption{An energy flow diagram between research questions (RQ1-RQ4, as defined in \tabref{tab:questions}) and bibliometric techniques (described in \tabref{tab:tools}).}
    \label{fig:flow-questions}
\end{figure}

\textcolor{black}{
In fact, we can benefit from this last mentioned diagram (also known as alluvial diagram \cite{Rosvall2010}) to illustrate the design and purpose of these analyses stages of the workflow. This is shown in \figref{fig:flow-questions}. Recall that RQ1 focuses on dynamics, while RQ2 and RQ3 refer to structures (knowledge and social, respectively). The left side of the diagram shows how each technique contributes some of the insights that help solve each of the research questions (note that some techniques can contribute to more than one question). The right-hand side, in turn, shows how the synthesis of findings from descriptive, trend, conceptual, social, and intellectual analysis ultimately adds to the critical wide-picture reflection of RQ4. The case study reported in Section \ref{sec:cases} provides a detailed discussion of the actual realisation of these flows.
}

\begin{figure}[h!]
    \centering
    \includegraphics[scale=.42]{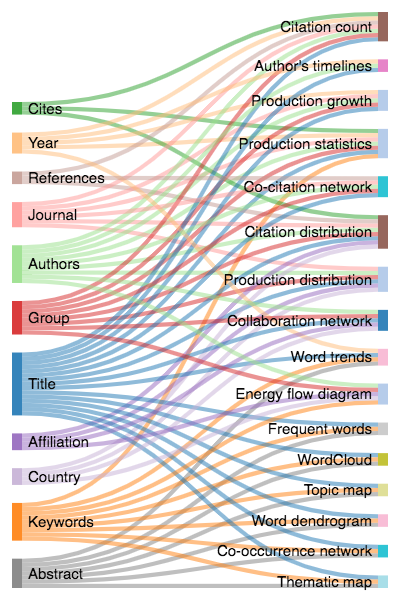}
    \caption{An energy flow diagram depicting the associations between metadata (shown in  \figref{fig:metadata}) and bibliometric techniques (described in \tabref{tab:tools}).}
    \label{fig:flow-metadata}
\end{figure}

For completeness sake, we also outline  the relations between the metadata scheme of the collected dataset (see \figref{fig:metadata}) and the set of bibliometric tools used in the analyses (see \tabref{tab:tools}). This is shown in the flow energy diagram of \figref{fig:flow-metadata}. There, it is observed that the main fields are Title, Author, Keywords and Summary, contributing predominantly to the fulfilment of most of the analyses. Colours in the flow diagrams are meant to enhance readability of energy transformations; they do not convey specific meaning.

\section{Case study}\label{sec:cases}

In order to demonstrate the application of the the described workflow, we conducted a study on two particular Master’s degrees from the School of Engineering, Universidad Distrital Francisco  José  de  Caldas  in  Bogotá,  Colombia:  the MSc. in Industrial Engineering and the MSc. in Information Sciences.  In this section we report the results of the study on the first, while in the interest of space, the results of the second are annexed to the Supplementary section. Besides, we have developed a companion web-based dashboard where all of these and more results can be browsed interactively (visit: \textcolor{blue}{\url{https://srojas.shinyapps.io/shinymasters/}}). For the sake of completeness, further discussion on the interpretation of the results will also be provided below, in light of each question. 

We note that for some specific analyses the results contain terms in Spanish, as this is the original writing language of these dissertations; this certainly does not limits the scope of application of the approach to analysing dissertations written in other languages, as long as the text cleaning lists are customised for that purpose. In this sense and in the interest of reproducibility, the datasets, R scripts and lists for text cleaning used in this study have been made publicly available (visit: \textcolor{blue}{\url{https://github.com/sargaleano/bibliomasters/}}).  

So, first of all, we set the observation period at 2010-2020, in order to analyze the entire last decade of program activity. Second, we extracted the metadata of the dissertations completed during that period from the institutional archive (\textcolor{blue}{\url{http://repository.udistrital.edu.co/}}). 

\subsection{Research landscape of the MSc. in Industrial Engineering, UDFJC.}

This programme was established on 2004 with a focus on the areas of Quality Assurance,
Operations Research and Statistics, and Occupational Health. In 2014, a new direction was given to the programme with emphasis on the areas of Logistics Systems, and including new lines of research on Organisation Management and Computational Intelligence for Business.
The dataset used to conduct the analysis, consisted of the metadata of the dissertations carried out during the 2010-2020 decade, compiled according to the guidelines provided in \secref{sec:dataset}. We will call it the \textit{MIE} dataset. Next, we will report the results of each stage in the proposed workflow.

\medskip
\noindent\textbf{\emph{RQ1. Production dynamics}}

\tabref{tab:statistics-MIE} summarises some descriptive statistics of the MIE dataset. Regarding the dynamics of production, a total of 143 dissertations were completed during the observation window. The average number of citations per document is relatively low (0.24 cites/document), compared to slightly higher averages found in the field of engineering  \cite{Kousha2019, Kousha2020}. The number of keywords per document is around 3.4 (486/143), a typical value. In contrast, the average number of completed dissertations per year is 13.0 (143/11 years), a low rate considering that roughly twice as many students enroll yearly in this programme. 

\begin{table}[b]
    \footnotesize
    \centering
    \setlength{\tabcolsep}{.2cm}
    \def\arraystretch{1.0}

    \begin{tabular}{|l|c|l|c|}
    \hline
    \multicolumn{2}{|c|}{\textbf{Dynamics}}     & \multicolumn{2}{c|}{\textbf{Structure}} \\ \hline
    Timespan                        & 2010-2020 & Authors                       & 188     \\ \hline
    Documents                       & 143       & Author appearances            & 295     \\ \hline
    Avg. citations per document     & 0.24      & Single-authored documents     & 4       \\ \hline
    Avg. citations per year per doc & 0.03      & Authors per document          & 1.31    \\ \hline
    Author's keywords               & 486       & Co-authors per documents      & 2.06    \\ \hline
    Unigram keywords                & 523       & Collaboration Index           & 1.32    \\ \hline
    Avg. dissertations per year     & 13.0      & References*                    & 3690  \\ \hline
    \multicolumn{4}{r}{*\scriptsize{References were only available since 2016 (46 documents)}} 
    \end{tabular}
    
        \caption{Bibliometric statistics for the MIE dataset.}
        \label{tab:statistics-MIE}
\end{table}

Incidentally, some of this statistics give us a glimpse of the structure of research production. Specifically, we found a total of 188 different authors. Notice that we assume that this number comprises the students who were actually the authors of the dissertation document together with their supervisor(s); thus, we reason that it corresponds to 143 students plus 45 faculty members. The number of author appearances is 295, which gives a ratio of 2.06 co-authors per document, meaning that a few students have had more than one supervisor. Only 4 documents are from a  single author (less than 3\%, possibly records that lack information about their supervisors). 

Additionally, the average number of unique authors per document is 1.31 (188/143), indicating that each faculty member must have supervised many dissertations. Moreover, the collaboration index (that is, the average number of authors in documents by multiple authors \cite{Elango2012}), yields a similar value of 1.32 (184/139), because all records are counted except those 4 missing supervisor information (139 out of 143 in total). This is a coherent value given that a dissertation is typically the result of a collaborative effort. Lastly, the ratio of references per document is 80.2 (3690/46) which is comparable with averages reported elsewhere for engineering programmes \cite{Kelly2015}. 

\begin{figure}
     \centering
     \begin{subfigure}[b]{0.45\textwidth}
         \centering
         \includegraphics[scale=.15]{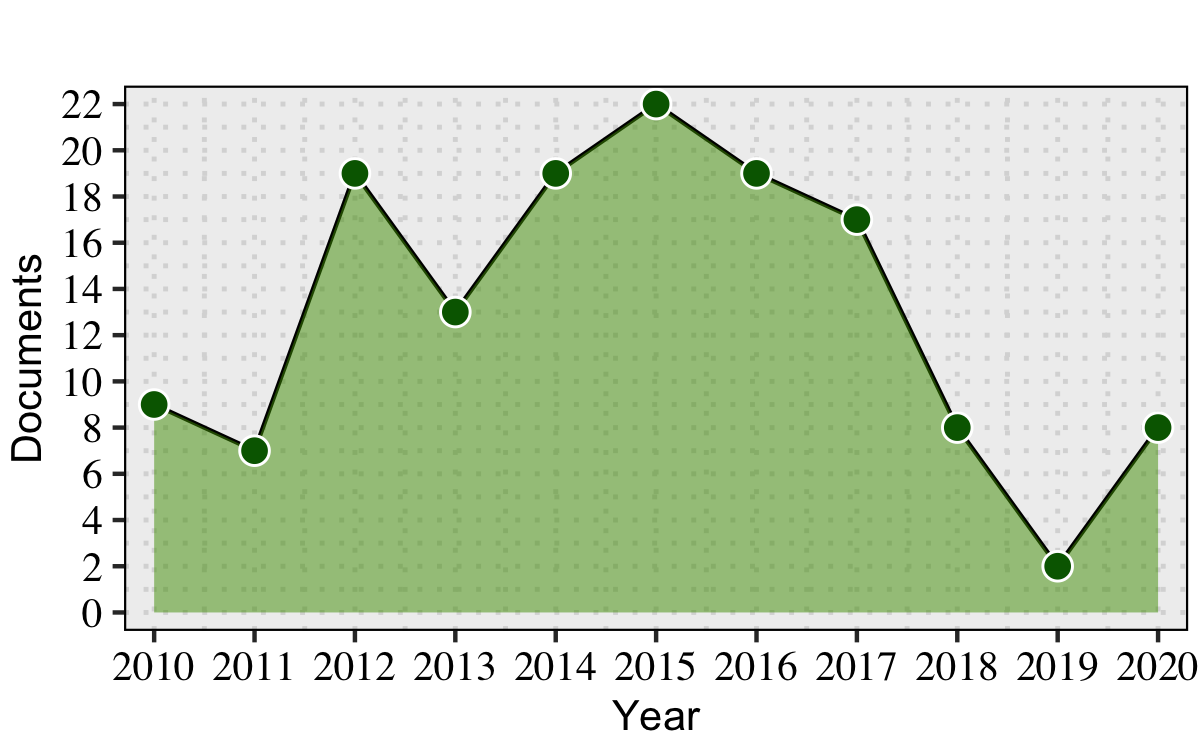}
         \caption{Production growth}
         \label{iind:production-growth}
     \end{subfigure}
     \hfill
     \begin{subfigure}[b]{0.45\textwidth}
         \centering
         \includegraphics[scale=.15]{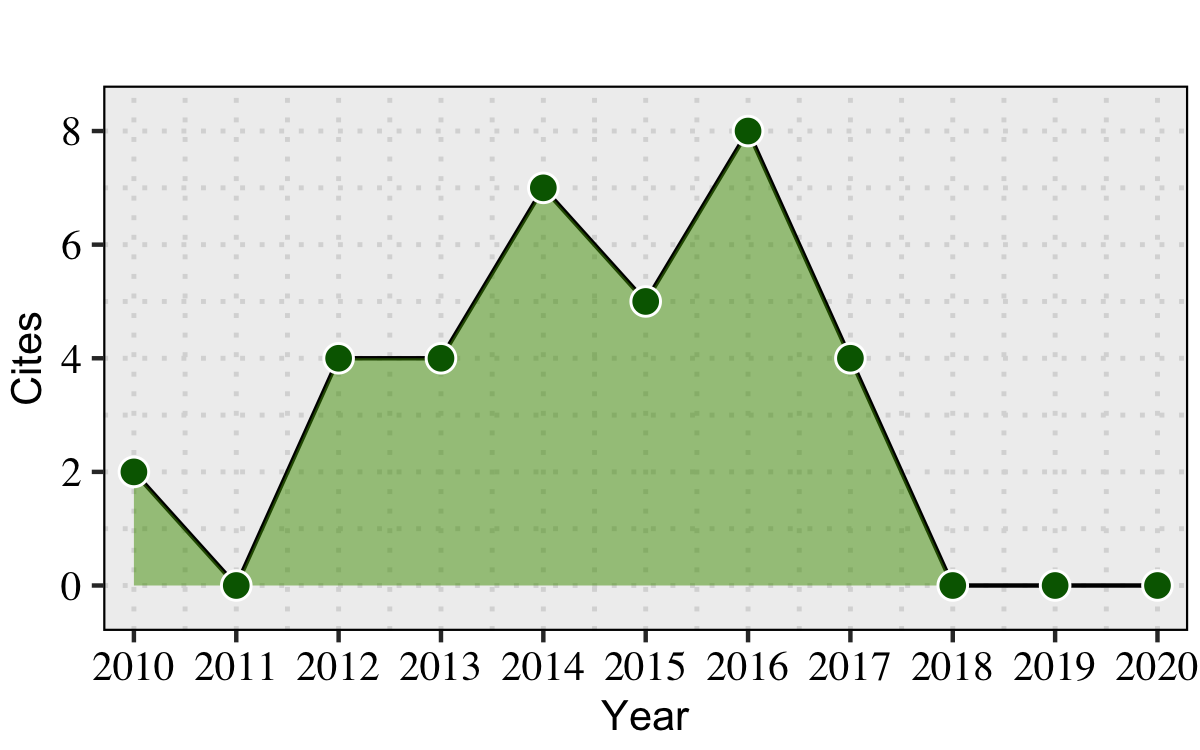}
         \caption{Citation count}
         \label{iind:citation-count}
     \end{subfigure}
     \hfill \\
     
     \begin{subfigure}[b]{0.45\textwidth}
         \centering
        \includegraphics[scale=.15]{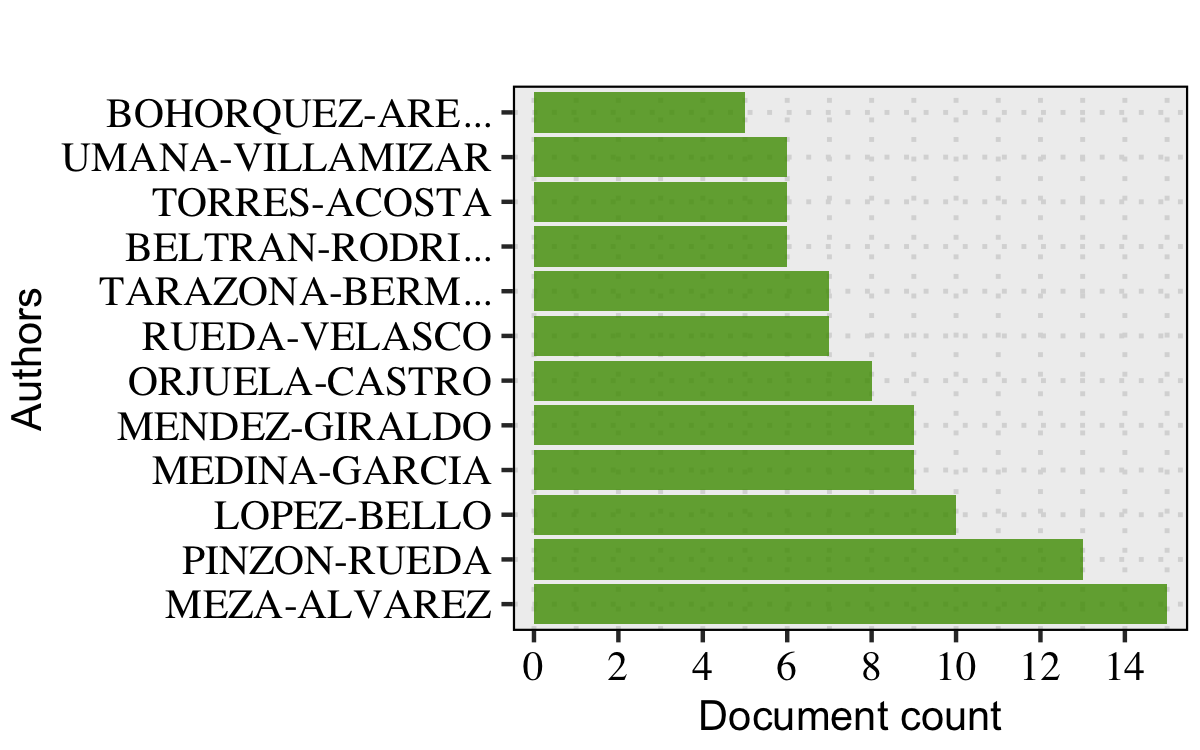}
        \caption{Production distribution (authors)}
        \label{iind:production-distribution-authors}
     \end{subfigure}
     \hfill
     \begin{subfigure}[b]{0.45\textwidth}
         \centering
         \includegraphics[scale=.15]{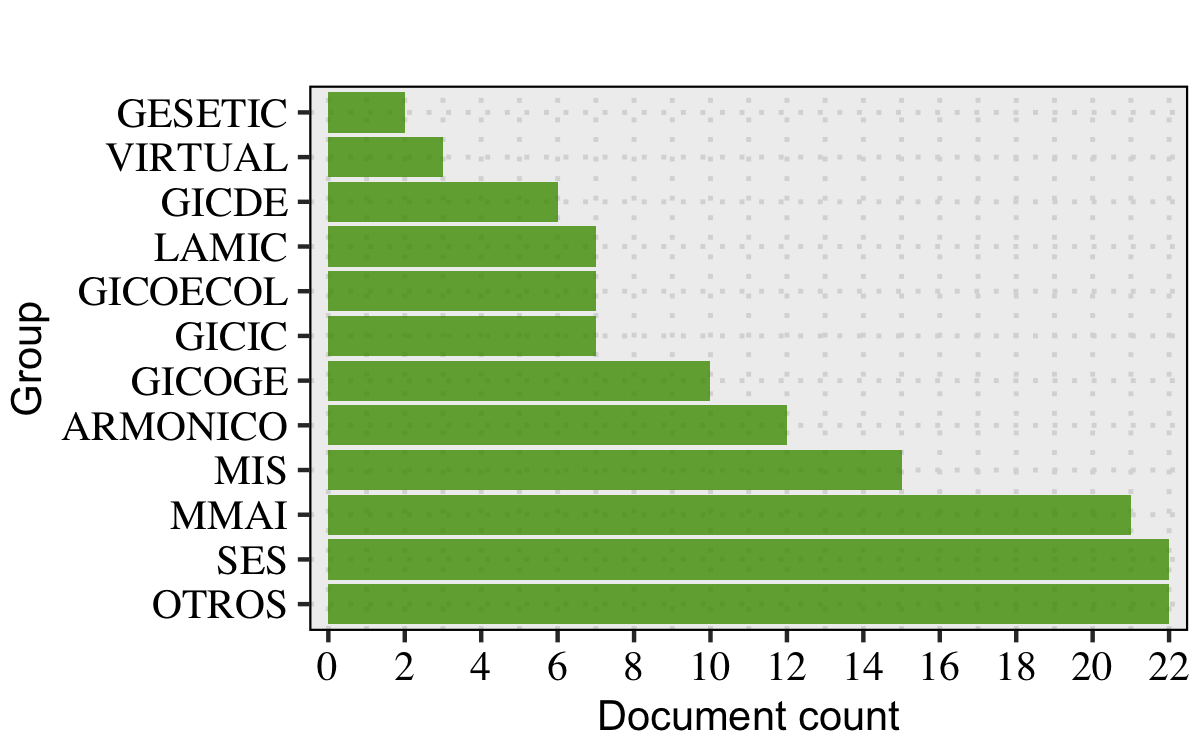}
         \caption{Production distribution (groups)}
         \label{iind:production-distribution-groups}
     \end{subfigure}
     \hfill \\
          
     \begin{subfigure}[b]{0.45\textwidth}
         \centering
        \includegraphics[scale=.15]{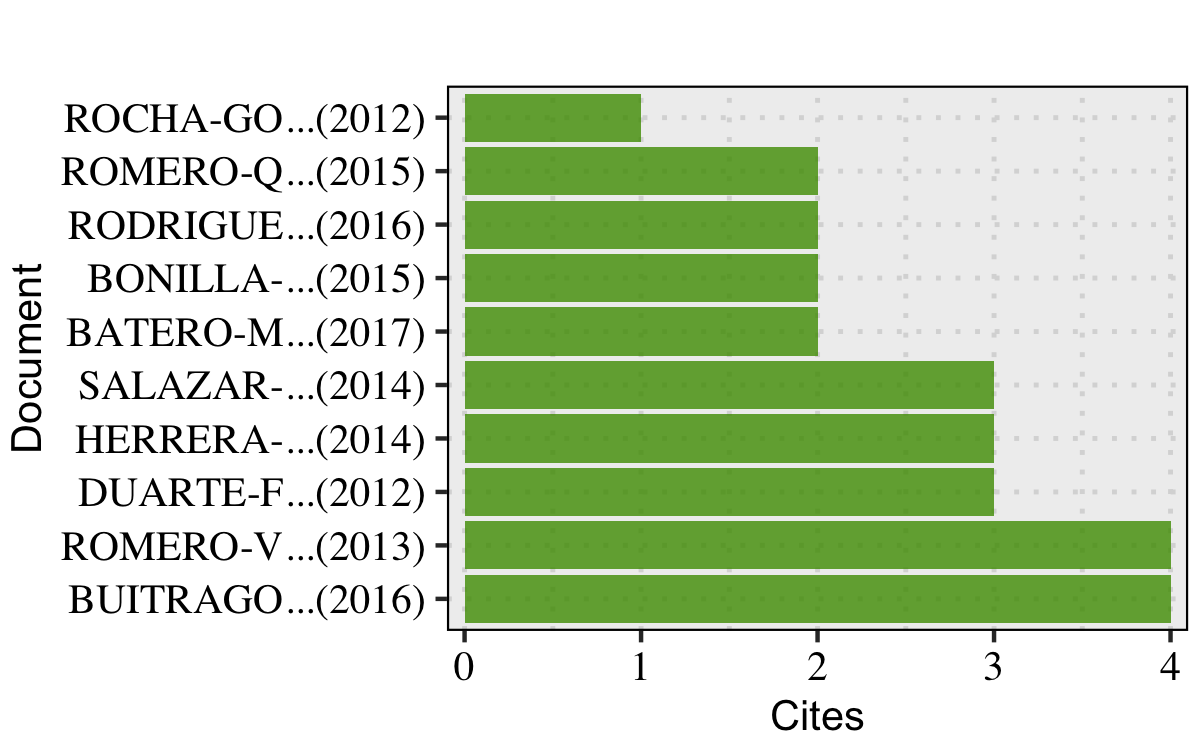}
        \caption{Citation distribution}
        \label{iind:citation-distribution}
     \end{subfigure}
     \hfill 
     \begin{subfigure}[b]{0.45\textwidth}
         \centering
         \includegraphics[scale=.15]{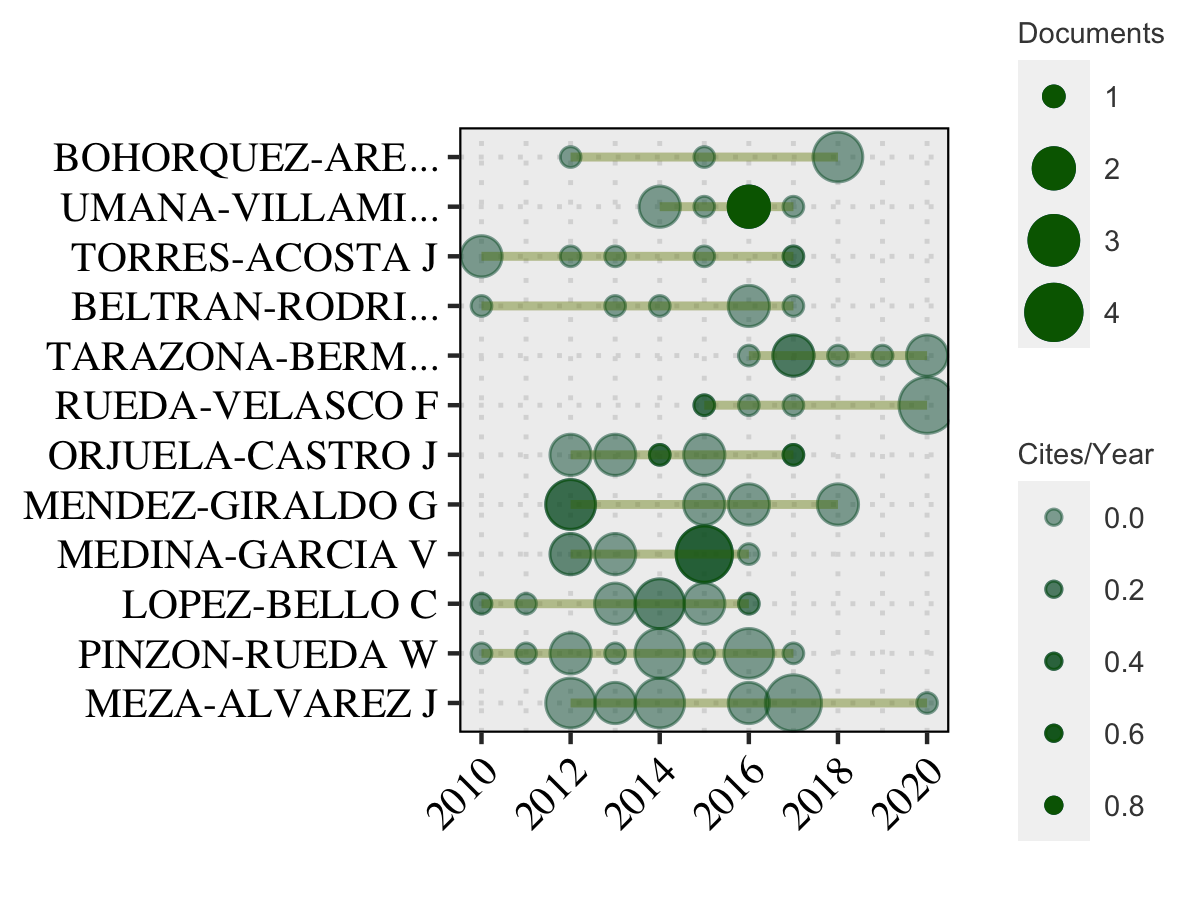}
         \caption{Author's timelines}
         \label{iind:authors-timelines}
     \end{subfigure}
     \hfill \\
     
     \begin{subfigure}[b]{0.45\textwidth}
         \centering
         \includegraphics[scale=.15]{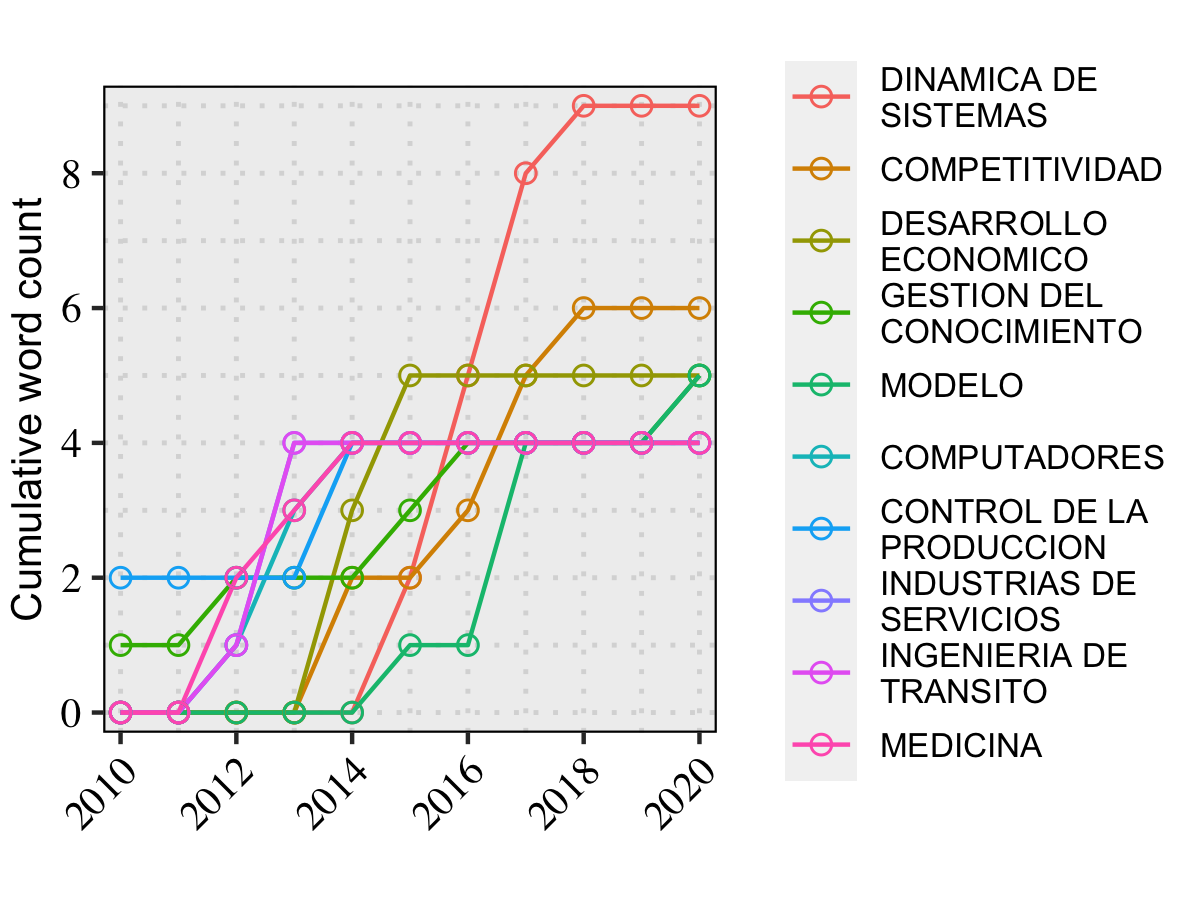}
         \caption{Word trends (keywords)}
         \label{iind:word-trends-keywords}
     \end{subfigure}
     \hfill
     \begin{subfigure}[b]{0.45\textwidth}
         \centering
         \includegraphics[scale=.15]{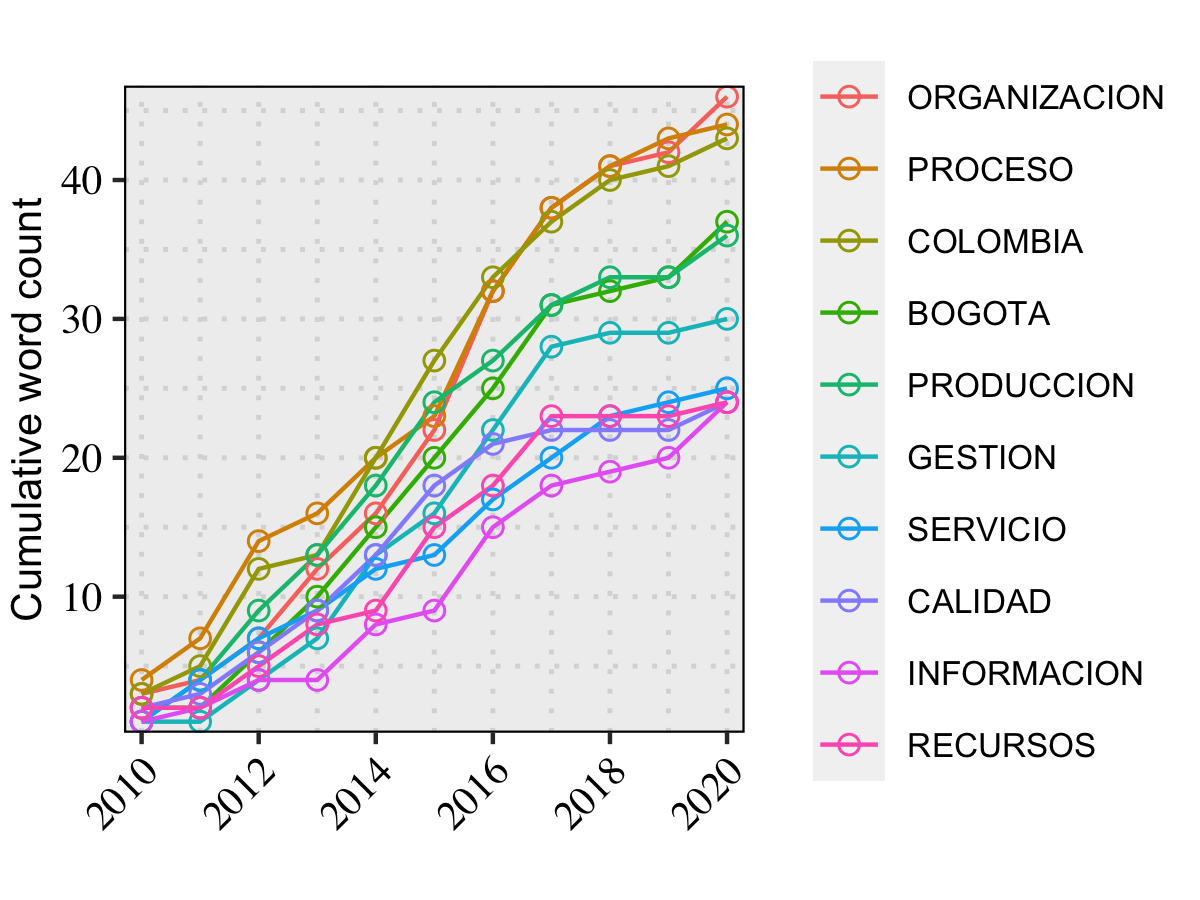}
         \caption{Word trends (abstracts)}
         \label{iind:word-trends-abstract}
     \end{subfigure}
     \hfill \\     
     
        \caption{Results of the RQ1 analysis (production dynamics) for the MIE dataset.}
        \label{fig:RQ1-MIE}
\end{figure}

The results of the additional analysis associated to RQ1 (see \figref{fig:flow-questions}) are reported in \figref{fig:RQ1-MIE}. \figref{iind:production-growth} shows two production peaks in 2012 with 19 dissertations, and in 2015 with 22 dissertations. The remaining years exhibited smaller numbers; a downward trend is observed during 2017-2019, which may suggest that students found it difficult to finish their dissertations during this interval (although an upturn in production is visible in 2020). Now, regarding citation dynamics (\figref{iind:citation-count}), dissertations completed in 2016 accumulated the highest number of citations (8); the overall curve shows a sawtooth pattern, but it is noticeable that no cites has been accrued by dissertations from 2018 to 2020, probably because it is too early in their maturity cycle.

On the other hand, \figref{iind:production-distribution-authors} shows the distribution of the 12 most prolific supervisors (considered co-author), adding up to 71\% of the documents in the dataset (101/143). Similarly, the distribution of research group affiliations is shown in \figref{iind:production-distribution-groups}; the most productive being \textit{SES} (Expert Systems and Simulation) and \textit{MMAI} (Mathematical Models Applied to Industry), representing 30\% (43/143) of the total number of dissertations between them. 

The analysis of the distribution of citations by individual document is shown in \figref{iind:citation-distribution}, where two dissertations appear as the most cited, each one with 4 citations: (Buitrago, 2016), ``\textit{Marco Conceptual del Conocimiento y el Aprendizaje Organizacional, del Enfoque Clasico al Enfoque de los Sistemas Adaptativos Complejos}'' and (Romero, 2013), ``\textit{Diseno de un Modelo de Controlador Flexible para un Sistema Integrado de Transporte que Permita Superar las Deficiencias Actuales en Captura de Datos e Intercambio entre Sistemas Heterogeneos}''; these dissertations focus on complex adaptive systems and control of heterogeneous transport systems, respectively. 

Another view of the production dynamics can be seen in the individual timelines of the most prolific authors (\figref{iind:authors-timelines}), where contribution size (document count) and contribution impact (citations per year) are plotted on an annual basis. This plot is useful for analysing the activity patterns of supervisors over time. It can be seen that the activity of the bulk of the group of supervisors has somewhat stagnated since 2017, with only four very active in the last 3 years: \textit{Bohorquez-Arevalo} and \textit{Mendez-Giraldo} in 2018, and \textit{Tarazona-Bermudez} and \textit{Rueda-Velazco} in 2020.  Similarly, this timeline analysis can be applied to the contributions of the research groups, as illustrated in the supplementary \figref{fig:timelines}, where a similar pattern is visible.

From a different angle, the evolution of word trends can provide an interesting picture of changes in the topics covered by the programme's dissertations over time. \figref{iind:word-trends-keywords} displays curves that describe the use of the author's keywords as a cumulative count of terms per year; there, System Dynamics (\textit{Dinamica de Sistemas}) is the fastest growing keyword, being the most used as of 2020 (with 9 dissertations mentioning), despite not having been used at all before 2014. It is followed by Business Competitiveness (\textit{Competitividad}) with 6 mentions as of 2020, rising from 2013 on. 

Word trend analysis can also be performed on terms extracted from the abstracts of the dissertations. As a result (see  \figref{iind:word-trends-abstract}), we find that the terms Business (\textit{Organizacion}) and Process (\textit{Proceso}) have been the most widely used during the observation window with nearly 45 counts each. These are followed by \textit{Colombia} and \textit{Bogota}, a reasonable result considering that this is the public university of the Bogotá District, making evident its immediate geographic area of influence.

\noindent\textbf{\emph{RQ2. Conceptual structures}}

\begin{figure}[t!]
     \centering

     \begin{subfigure}[b]{0.45\textwidth}
        \hspace{-1cm}
        \includegraphics[scale=.15]{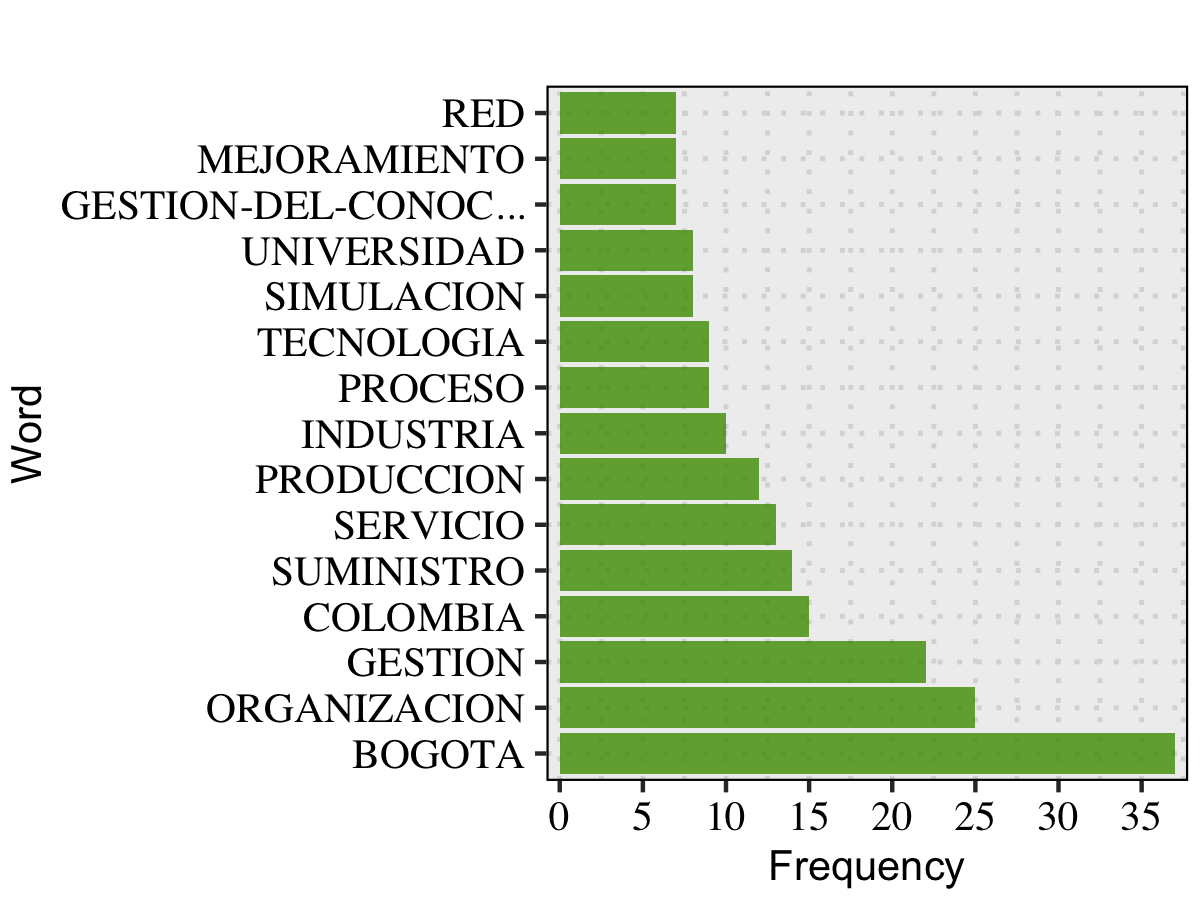}
        \caption{Frequent words (titles)}
        \label{iind:frequent-words}
     \end{subfigure}
     \hfill
     \begin{subfigure}[b]{0.45\textwidth}
         \centering
        \includegraphics[scale=.25]{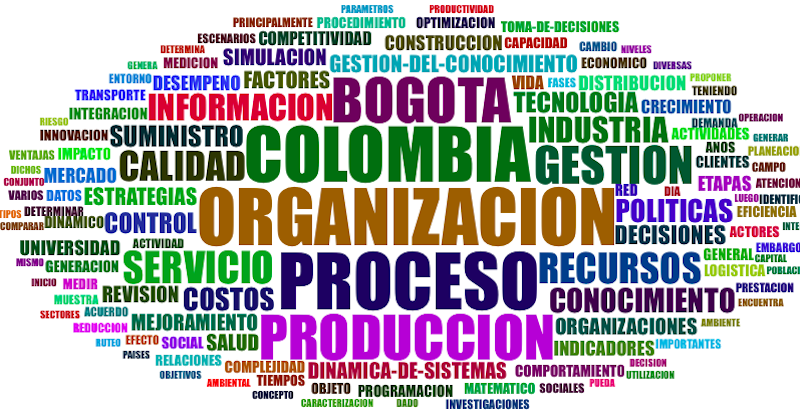}
        \caption{Word cloud (abstracts)}
        \label{iind:word-cloud}
     \end{subfigure}
     \hfill \\
     
     \begin{subfigure}[b]{0.45\textwidth}
         \centering
        \includegraphics[scale=.15]{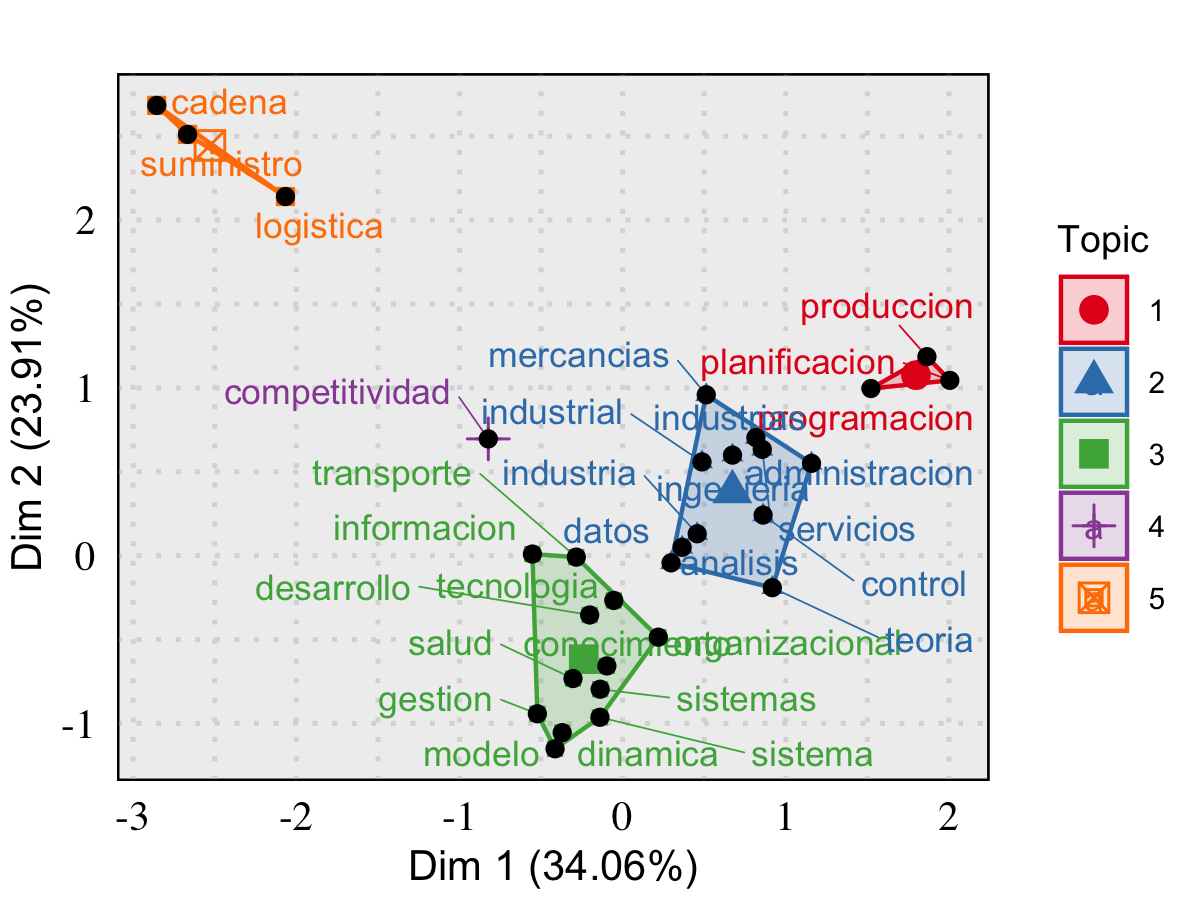}
        \caption{Topic map (keywords)}
        \label{iind:topic-map}
     \end{subfigure}
     \hfill
     \begin{subfigure}[b]{0.45\textwidth}
         \centering
        \includegraphics[scale=.15]{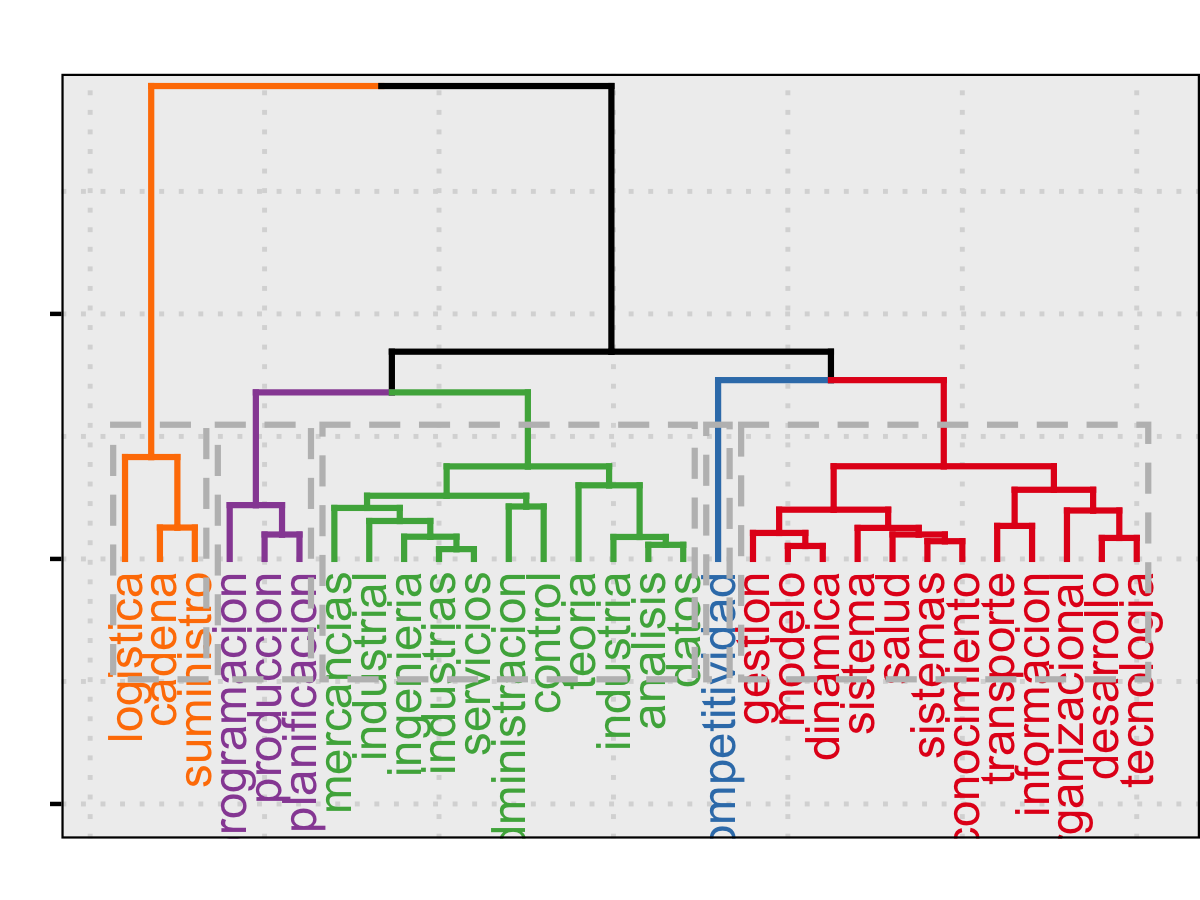}
        \caption{Word dendrogram (keywords)}
        \label{iind:word-dendrogram}
     \end{subfigure}
     \hfill \\
     
     \begin{subfigure}[b]{0.45\textwidth}
        \hspace{-1cm}
        \includegraphics[scale=.45]{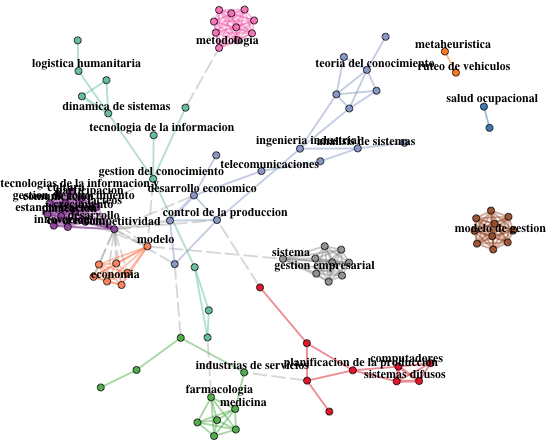}
        \caption{Co-occurrence network (keywords)}
        \label{iind:coocurrence-network}
     \end{subfigure}
     \hfill
     \begin{subfigure}[b]{0.45\textwidth}
         \centering
        \includegraphics[scale=.12]{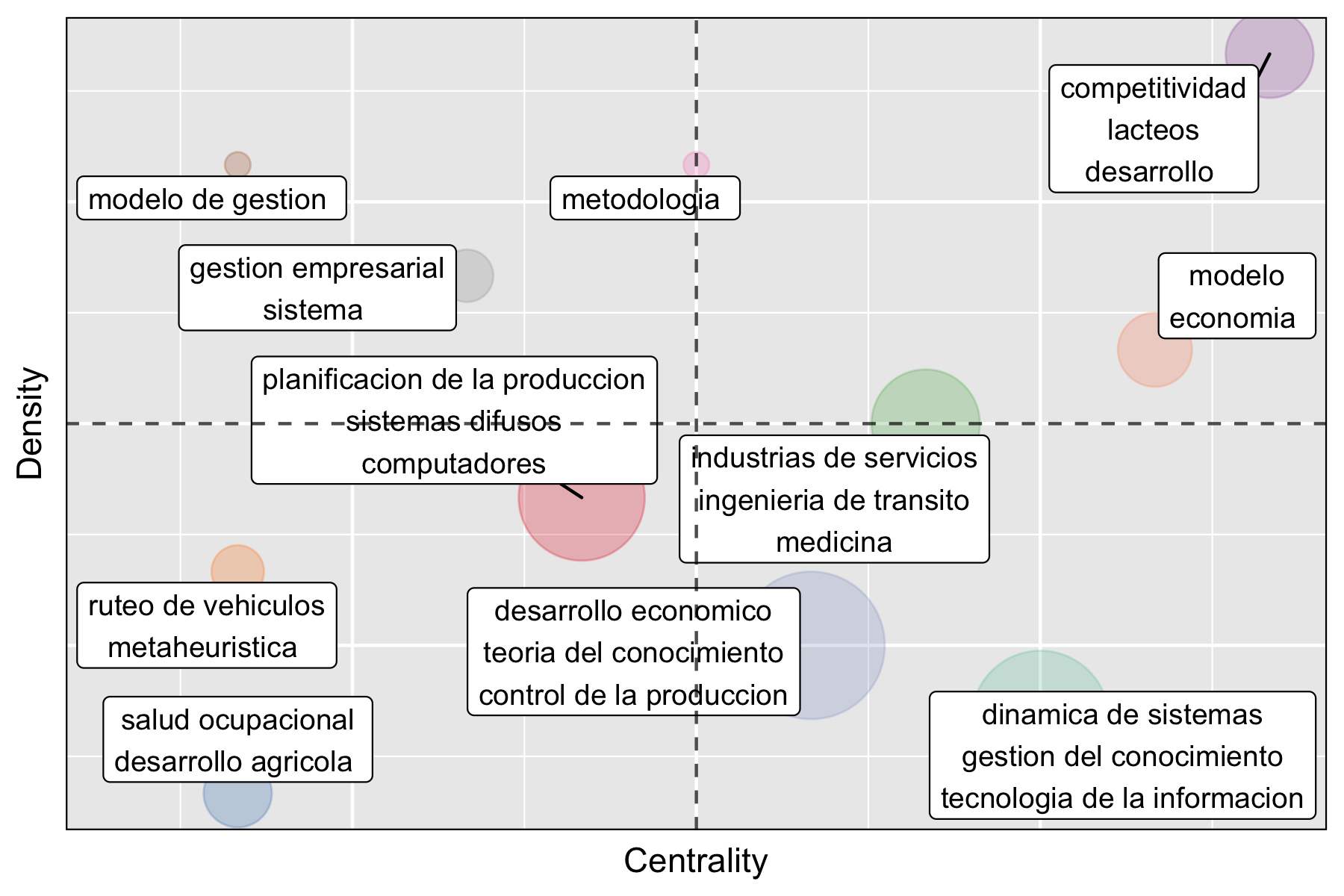}
        \caption{Thematic map (keywords)}
        \label{iind:thematic-map}
     \end{subfigure}
     
        \caption{Results of the RQ2 analysis (conceptual structures) for the MIE dataset.}
        \label{fig:RQ2-MIE}
\end{figure}

The analyses carried out for this question aim to discover thematic areas, dominant and emerging topics, and strengths of the research groups and faculty affiliated with the Masters programme. The results of the MIE dataset are summarised in \figref{fig:RQ2-MIE}. First, the plot of the 15 most frequent words used in the titles during the entire observation window is shown in \figref{iind:frequent-words}; the term \textit{Bogota} appears in 25\% of them (36/143), supporting the case of the relevance of the programme to help expand the scope of industrial engineering applications for the capital city of Colombia. The other terms are related to pertinent concepts such as Business (\textit{Organizacion}), Management (\textit{Gestion}), Supply (\textit{Suministro}), Production (\textit{Produccion}), Industry (\textit{Industria}), Simulation (\textit{Simulacion}), etc.

Another appealing choice to visualise the most frequent terms is a word cloud plot, where frequency and relevance are displayed by the size and central location of the words within the cloud (colors are used for readability only). \figref{iind:word-cloud} shows the word cloud of the terms used in the corpus abstracts. The most prominent ones actually correspond to those shown in the frequency histogram mentioned above, but the word cloud allows a broader display of many more terms. 

Note that it is possible to obtain both frequency histograms and word clouds from the different text fields found in the metadata records, that is, title, abstract, author keywords, and unigram keywords. By contrasting the plots derived from these fields, the analyst may gain an enriched understanding of the trends and patterns found in relation to the most prominent descriptors, index terms and word categories used by authors in a particular observation period. For the sake of clarity,  \figref{fig:frequent-words} and \figref{fig:wordclouds} of the supplementary material section, illustrate this point.

We now turn to the topic map of \figref{iind:topic-map}, where groups of related concepts (``topics'') are shown  representing the knowledge structures most strongly developed by the examined dissertations. Topics are formed by grouping terms that are proximal, in the sense that they are treated together in a large proportion of documents in the dataset. There, the 5 most important  topics that emerge are: production planning (topic 1, red), industry services (topic 2, blue), business management (topic 3, green), competitiveness (topic 4, purple) and logistics (topic 5, orange). 

The visualisation of the topic map is obtained by processing the term-to-document occurrence matrix \cite{Aria2017, Batagelj2013} and applying a dimensionality reduction technique to obtain a 2D projection on the two dimensions embedding the widest variability (here we used the MCA algorithm \cite{Cuccurullo2016}). 

In \figref{iind:word-dendrogram} one can see an alternative view of the topic map, known as a dendrogram. In this representation, a hierarchical tree is built from the  associations found between proximal terms. Therefore, each topic correspond to the set of terms sharing a common ancestral branch in the tree. Different groupings can emerge as one navigates through the levels of the hierarchy; thus, a cutoff level must be chosen. In this case, we chose the cutoff that produced the same clusters as those in the topic map shown before (albeit with a distinct colour legend). One of the advantages of dendrograms is that they allow greater readability of the terms included in each topic; another is the ability to find more generic or more specialised topics as the analyst move the cutoff level up or down in the hierarchy.

Let us comment that again, both topic maps and dendrograms can be generated from the various text fields in the metadata, so visualising and comparing them can be useful to capture a broader picture of the knowledge structures that develop from the dataset. As an example, we report those plots in the supplementary section, \figref{fig:topic-maps} and \figref{fig:dendrograms}.

Another useful approach to discovering the underlying  conceptual structures of the programme's research landscape, is to analyse the co-occurrence of terms in subsets of documents to derive a network graph, such as that obtained from the author's keywords in the dataset (as seen in \figref{iind:coocurrence-network}). Here, mainstream concepts appear in the central area of the network, while the unconventional or highly specialized concepts will be placed on the periphery. For instance, the network in the figure shows as core concepts, Business management (\textit{gestion empresarial}), Production planning (\textit{control de la produccion}) or Knowledge management (\textit{gestion del conocimiento}), whilst System Dynamics (\textit{dinamica de sistemas}, Humanitarian logistics \textit{logistica humanitaria}), Metaheuristics  (\textit{metaheuristicas}) or Fuzzy Systems, (\textit{sistemas difusos}) appear as specialised concepts. 

In the co-occurrence network, the strength of the relationship is visualised as the intensity of the edges and the proximity of the vertices. As a result, it is also possible to identify clusters of concepts that indicate the formation of underlying topics; for example, the brown and green clusters in \figref{iind:coocurrence-network} correspond to the Medicine (\textit{medicina}) and Management models (\textit{modelos de gestion}) topics, respectively. Furthermore, these networks can also be generated from the other text fields in the metadata (e.g. see supplementary \figref{fig:cooccurrence}).

Along the same lines, an alternate representation of the conceptual structures that can be derived from the co-occurrence network is the thematic map. To do this, the topics of the network are projected onto a 2D map whose dimensions are centrality (relevance of a theme in the research field) and density (maturity on the  development of a theme). Therefore, the four quadrants of the map (counterclockwise) would represent motor themes (first quadrant), isolated but highly specialised themes (second), emerging themes (third) and fundamental themes (fourth); centrality and density are calculated from the co-occurrence of keywords network (see \cite{Cobo2011} for details). \figref{iind:thematic-map} illustrates the thematic map of the analysed dataset, where topics related to System Dynamics and Service Industries appear as fundamental themes,  Competitiveness and Economy appear as motor themes, while Fuzzy Systems and Metaheuristics as emerging themes. The companion thematic map generated from unigram keywords is included in the supplementary \figref{fig:thematic-maps}.

\begin{figure}[b]
    \centering
    \includegraphics[scale=.3]{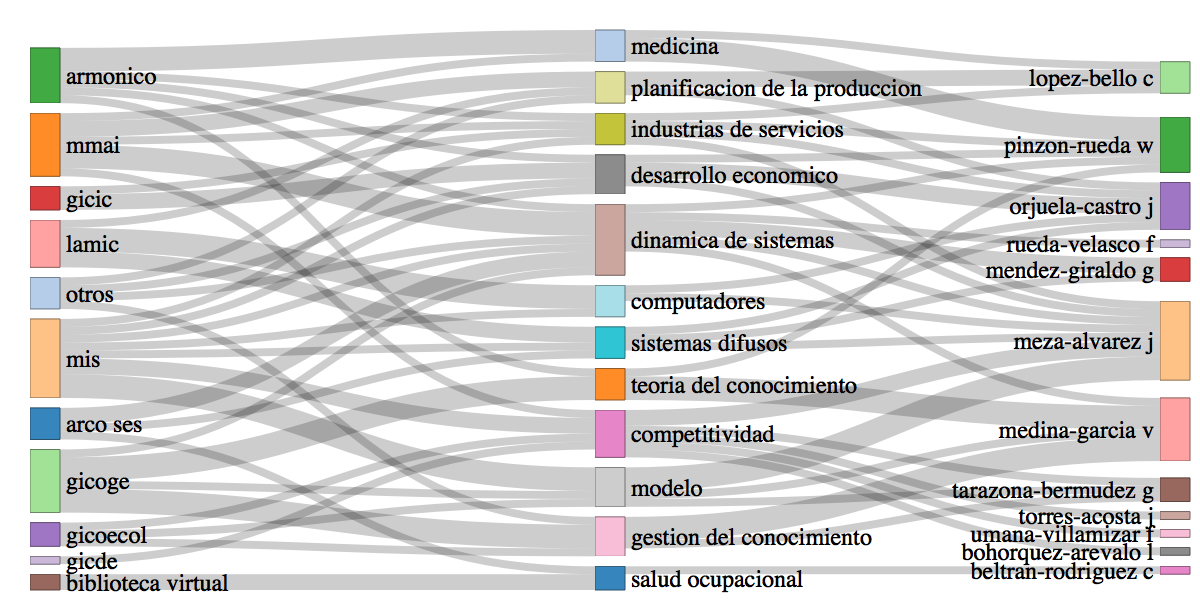}
    \caption{Energy flow through conceptual structures (MIE dataset).}
    \label{iind:energy-flow-conceptual}
\end{figure}

Lastly, to conclude the analysis of the conceptual structures, patterns of concept contributions from dissertations associated to groups or supervisors, throughout their author's keywords can be represented in an energy flow diagram (also known as alluvial diagrams \cite{Rosvall2010}), like the one in \figref{iind:energy-flow-conceptual}. Keywords are positioned in the middle of the flow, between the most prolific groups and authors; in this way it is possible to identify strengths of the groups, as well as the expertise of supervisors. The figure shows that, as before, System Dynamics and Competitiveness are two of the most dominant concepts, receiving contributions from many groups and supervisors. Additionally, the emerging or specialised topics identified in the previous analyses, can also be discovered here (e.g. Fuzzy Systems).

\medskip
\noindent\textbf{\emph{RQ3. Collaboration structures}}

\begin{figure}[t]
     \centering
     \begin{subfigure}[b]{0.45\textwidth}
        \hspace{-1.5cm}
        \includegraphics[scale=.3]{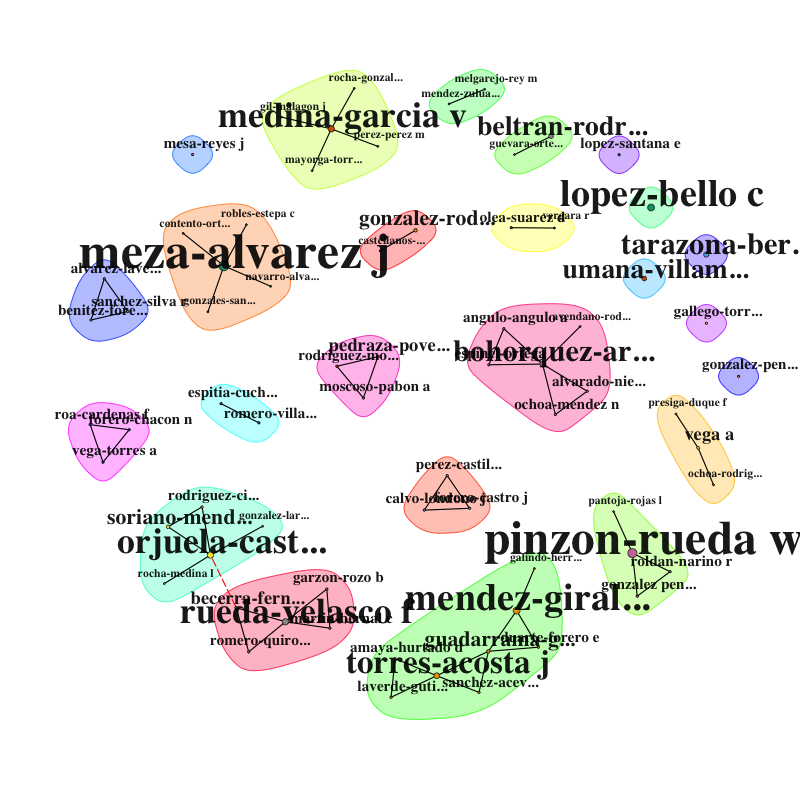}
        \caption{Authors collaboration network}
        \label{iind:collaboration-authors}
     \end{subfigure}
     \hfill
     \begin{subfigure}[b]{0.45\textwidth}
        \includegraphics[scale=.3]{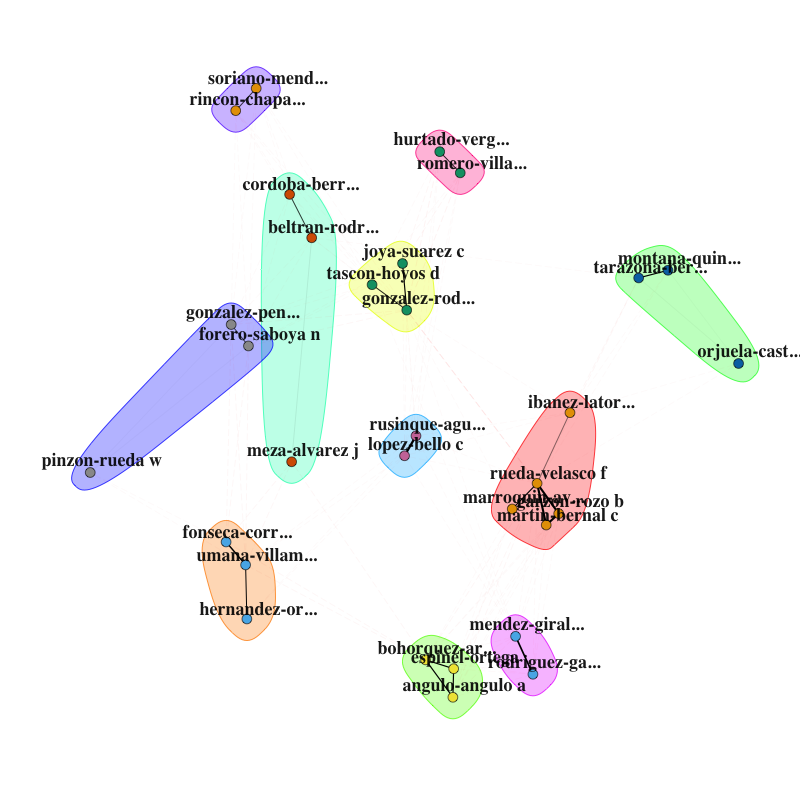}
        \caption{Authors coupling network}
        \label{iind:coupling-authors}
     \end{subfigure}

     \centering
     \begin{subfigure}[b]{0.45\textwidth}
        \hspace{-1.5cm}
        \includegraphics[scale=.3]{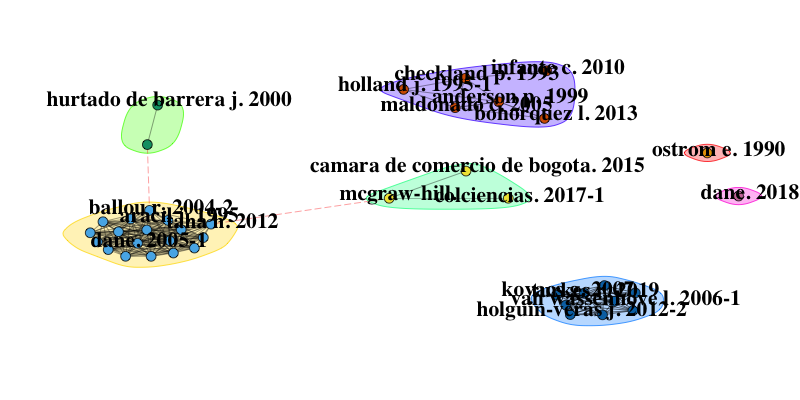}
        \caption{Co-citation references network}
        \label{iind:cocitation-references}
     \end{subfigure}
     \hfill
     \begin{subfigure}[b]{0.45\textwidth}
        \includegraphics[scale=.3]{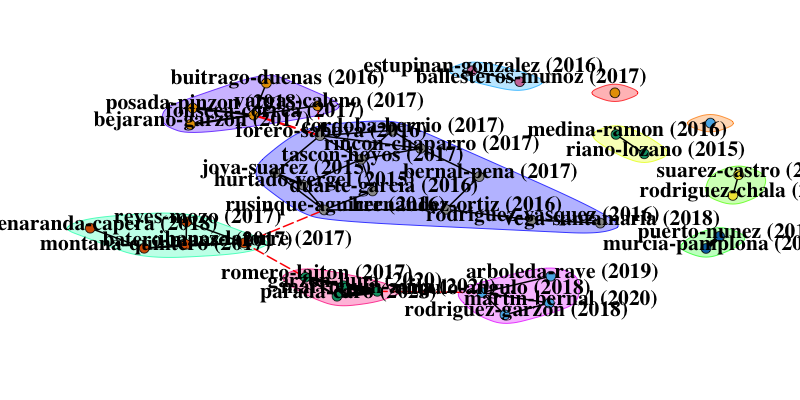}
        \caption{Manuscript coupling network}
        \label{iind:coupling-manuscripts}
     \end{subfigure}

        \caption{Results of the RQ3 analysis (collaboration structures) for the MIE dataset.}
        \label{fig:RQ3-MIE}
\end{figure}

The aim in this stage is to reveal the collaboration patterns implicitly evolved within the Master's programme, considering the social networks of authors and groups, and the intellectual networks of common references found among the dissertations. The results are shown in \figref{fig:RQ3-MIE}.\\[1cm]

Let us start discussing the network of authors collaboration shown in \figref{iind:collaboration-authors}, which is obtained by finding the co-occurrences in the list of authors of the dissertations. Since we assumed authors comprise the students and their supervisors, we can identify three different types of structures in the network. First, there are star-shaped clusters, in which a single supervisor (central node in the cluster) has collaborated with many students to produce several dissertations (see e.g. clusters with the central label \textit{meza-alvarez} or \textit{medina-garcia} in the figure); the number of dissertations is proportional to the size of the central label. 

Secondly, there are some triangle-shape clusters, which indicate that two supervisors collaborated with a student in his/her dissertation (e.g. the \textit{benitez+sanchez+alvarez} or the \textit{perez+calvo+castro} clusters). And thirdly, there are larger clusters that combine the previous two types, representing extended links of collaboration between several supervisors and students (e.g the \textit{orjuela+soriano+rueda} or \textit{mendez+torres+guadarrama} clusters). The latter suggests the formation of communities.

Now let us move on to the author's bibliographic coupling network of \figref{iind:coupling-authors}. In this network, two authors are connected if they have a common reference cited in the references list of their oeuvres included in the dataset \cite{Zhao2008}; in this case, the oeuvres would be individual dissertations for student authors, or the sets of supervised dissertations, for supervisors. As a result, the formation of several clusters of active authors sharing research interests can be discovered. 

Another perspective of intellectual structures, is given by the network of co-cited references, that is, the frequency with which two references are cited jointly across many manuscripts \cite{Small1973}. This network provides a glimpse of influential works in the literature that are being referenced in subsets of the analysed dataset. As a complement of the topic and thematic maps, this intellectual network may be useful to identify paradigms, or influential authors adopted by the Master's communities, as shown in \figref{iind:cocitation-references}.   

A closely related analysis of intellectual structures is the  manuscript coupling network (\figref{iind:coupling-manuscripts}). In this case, the connections arise when the dissertations refer to shared works in their bibliographies. Therefore, this network identifies dissertations that develop related themes using a common theoretical framework. That said, we note that the analysis can be extended to a variety of other type of bibliographic couplings, so as to discover further social or intellectual structures underlying the programme (see \cite{Qiu2014} or also an in-depth discussion in \cite{Batagelj2013}).

\begin{figure}[t]
    \centering
    \includegraphics[scale=.3]{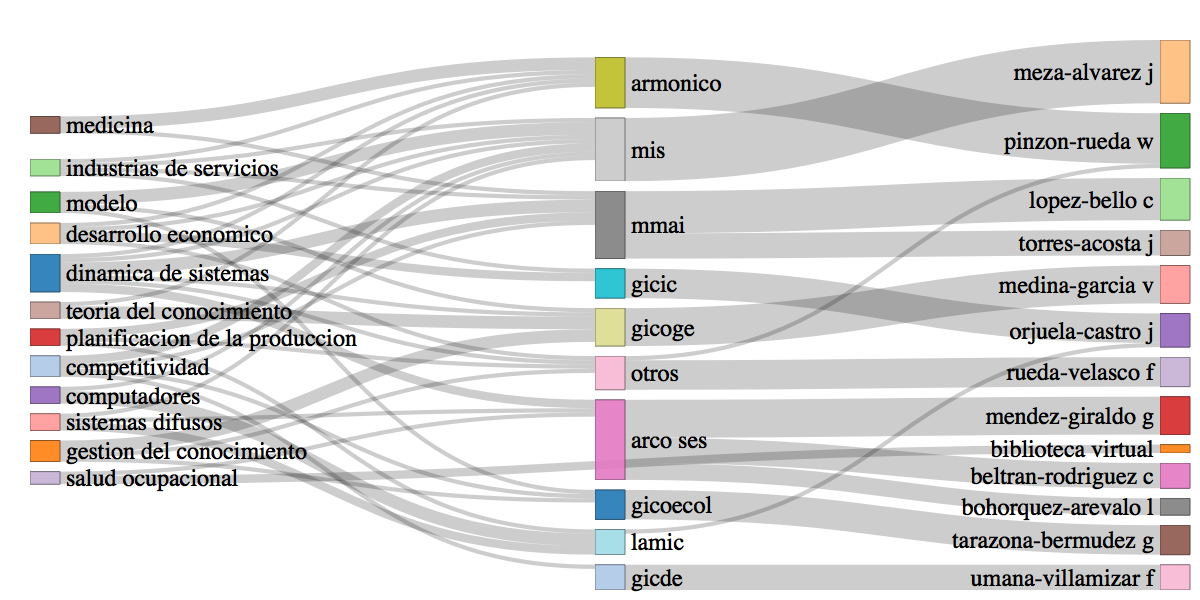}
    \caption{Energy flow through social structures (MIE dataset).}
    \label{iind:energy-flow-social}
\end{figure}

Lastly, we note that the aforementioned energy flow diagram can also be used to visualise  collaboration structures between groups and authors, such as in \figref{iind:energy-flow-social}. Here, the widths of the authors bands are proportional to the amount of supervised dissertations.

\medskip
\noindent\textbf{\emph{RQ4. Critical assessment of the Master's research landscape}}

The findings reported for RQ1, RQ2 and RQ3 in previous sections, provide a comprehensive view of the research landscape of the Masters' programme, which allow us to draw the following conclusions. During the 2010-2020 period, the MIE programme exhibited a moderate production output (average of 13.0 dissertations per year), considering that the student admission rate is nearly twice this number per year. Besides, the dynamics of production remained stable until 2016, but it has been decreasing from 2017-2020. It would be prudent to follow up with the authors to verify the difficulties encountered that delay the research plan of the dissertations started in that subperiod. 

On the other hand, supervision has been carried out by 45 professors, although a biased distribution was found towards 12 supervisors, accounting for around 71\% of the thesis production during the observation period. Given that only four faculty members were active as supervisors of completed dissertations in the last 3 years, it would be cautious to secure additional support (in funding or dedication time) to promote the willingness to assume supervision duties.

Now, the research strengths of the programme are mainly related to the following driving themes: production planning, system dynamics and industry competitiveness. Nonetheless, there are emerging topics gaining momentum, such as fuzzy systems and metaheuristics. In a way, these topics can be associated to the new focus given to the programme in 2014, emphasising in Logistics (that can be related to production planning), Organisation Management (related to system dynamics and industry competitiveness) and Business Intelligence (related to fuzzy systems and metaheuristics). 

Regarding citation impact, an average 0.24 cites per manuscript is relatively low compared to other international programmes; curiously enough, the two most cited dissertations addressed the topics of complex systems and transport systems, which are not closely related to those dominant or emerging topics mentioned above. Thus, on the one hand, it would be interesting to reflect on the contribution of the dominant and emerging thematic areas towards the strategic goals of the programme proposed for the short- and medium-term. And, on the other hand, it is recommended to promote a wider visibility of previous works in the incoming students, to facilitate growth of thematic areas addressed in past dissertations.

Additionally, it is worth noting the programme is deeply motivated to propose industrial engineering applications in the context of the capital city, as nearly a quarter of the dissertations included the term ``Bogotá'' in their titles or index terms. This is in line with the historical roots and closeness that the city maintains with the university, which incidentally is also its main sponsor.

In terms of collaboration structures, the programme has developed a few communities of multiple faculty researchers working on related subjects, although most of the collaboration is accomplished as isolated clusters of supervisors working alone with their students. This suggests that intragroup collaborations are rare, despite the fact that most groups consists of several faculty members rather than a small number of one-person groups. Therefore, initiatives to promote information-sharing and co-working, including internal seminars and workshops, would be strongly recommended. 

\section{Conclusion}
The workflow described in this paper leverages a variety of complementary bibliometric facets (descriptive, trends, conceptual, social and intellectual analyses) to assess the research output landscape of a MSc. programme, with respect to the structure and dynamics patterns emerging during an observation window. The insights gained from each analysis are aggregated to obtain a comprehensive picture of such landscape, as demonstrated by the reported cases. 

If carried out regularly, the workflow can be used to track the evolution of the programme's behaviour over time, providing decision-makers with actionable insights to guide the short, medium and large--term strategic planning. Thus, it may also be advantageous to perform comparative studies with similar programmes from different institutions, or to help measure its level of maturity or academic quality achievements. 

The multifaceted nature of our approach is in accordance with the increasingly adopted stance of a multidimensional understanding of the scientific impact and quality of research output, in contrast to a citations-focused appraisal (see \cite{Aksnes2019} and references within). We advocate that these multidimensional assessments provide a more critical and  comprehensive overviews of the strengths and weaknesses of the programme, benefiting the stakeholders, whether being those defining the orientations inside the programme, or those who review it externally from the funding agencies or government research policy agencies. 

Finally, the workflow can be considered as a framework for analysing master's profiles for a variety of products, such as papers, grey literature, or software developed by the groups and researchers associated with it. Likewise, it can be extended to other existing or novel bibliometrics techniques, or to different bibliometric software tools. In fact, we anticipate the approach can be applied also to other academic units, such as PhD. courses, research labs and institutes, or even entire graduate schools; how to establish the length of the observation window and the frequency of application depending on the discipline, the purposes and the unit of the study, are interesting questions to address in future work.

\section*{Acknowledgements}
\textcolor{black}{
The authors are grateful to Lindsay \'{A}lvarez from the School of Engineering of Universidad Distrital Francisco Jos\'{e} de Caldas, for her valuable comments and discussions that helped shape the tone of the manuscript. They also would like to thank Carolina Suárez from the same school, for her careful review and suggestions that helped clarify some aspects of the text.  
}
\bibliographystyle{plain}
\bibliography{biblio}

\clearpage

\renewcommand{\thepage}{S\arabic{page}}  
\renewcommand{\thesection}{S\arabic{section}}   
\renewcommand{\thetable}{S\arabic{table}}   
\renewcommand{\thefigure}{S\arabic{figure}}

\setcounter{section}{0}
\setcounter{table}{0}
\setcounter{figure}{0}

\section*{Supplementary material}

\section{Datasets and code repository}
A companion interactive web-based dashboard to perform the analyses is available at: 

\textcolor{blue}{\url{https://srojas.shinyapps.io/shinymasters/}}

\medskip
\noindent A public repository with datasets, R scripts and lists for text preprocessing that were used in this study, is available at: 

\textcolor{blue}{\url{https://github.com/sargaleano/bibliomasters/}}

\section{Supplementary analysis on the MIE dataset}

\subsection{Individual timelines}
\begin{figure}[H]
     \begin{subfigure}[b]{0.5\textwidth}
        \includegraphics[scale=.18]{IIND/authors-timeline.png}
        \caption{Field: Author}
     \end{subfigure}
     \hfill
     \begin{subfigure}[b]{0.5\textwidth}
        \includegraphics[scale=.18]{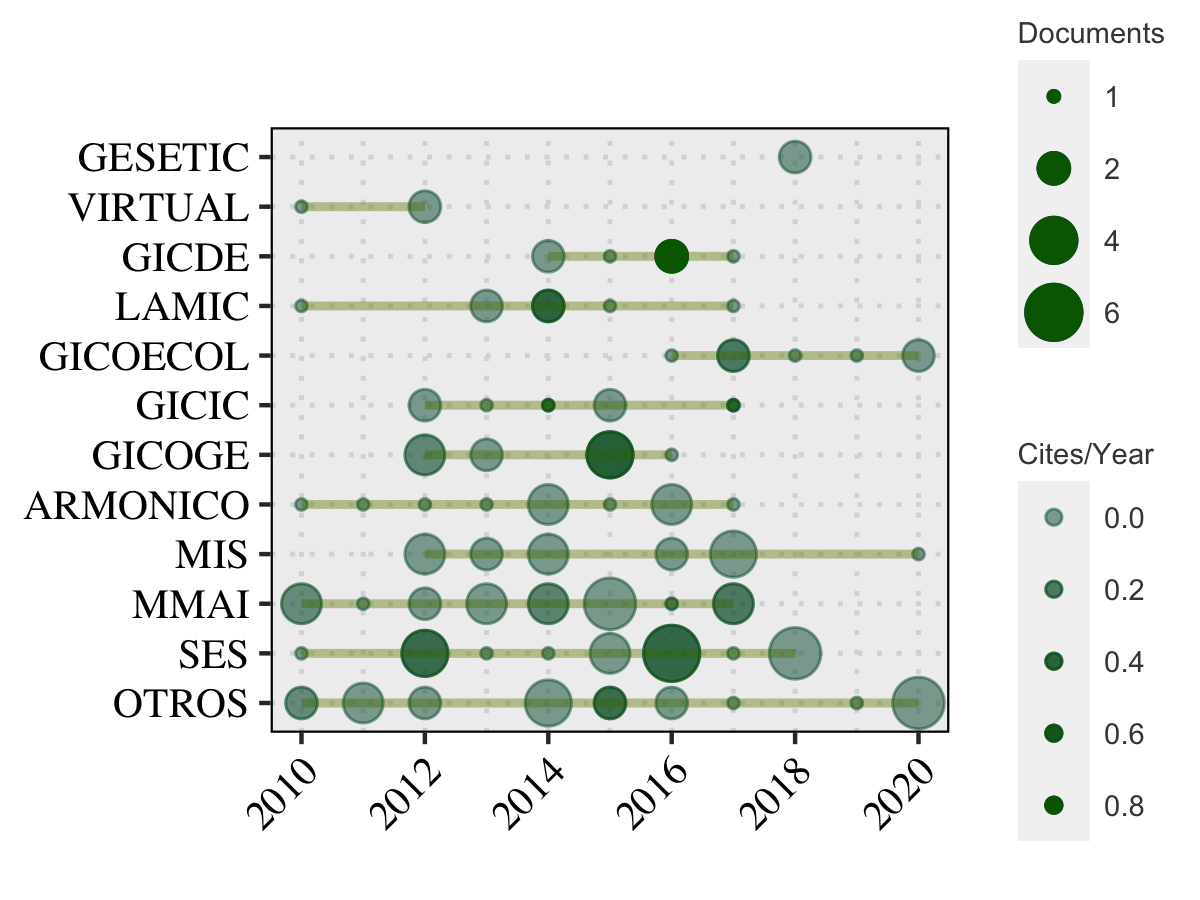}
        \caption{Field: Group}
     \end{subfigure}
     
        \caption{Individual timelines using different fields for the MIE dataset.}
        \label{fig:timelines}
\end{figure}

\subsection{Frequent words}
\begin{figure}[H]
     \begin{subfigure}[b]{0.5\textwidth}
        \includegraphics[scale=.18]{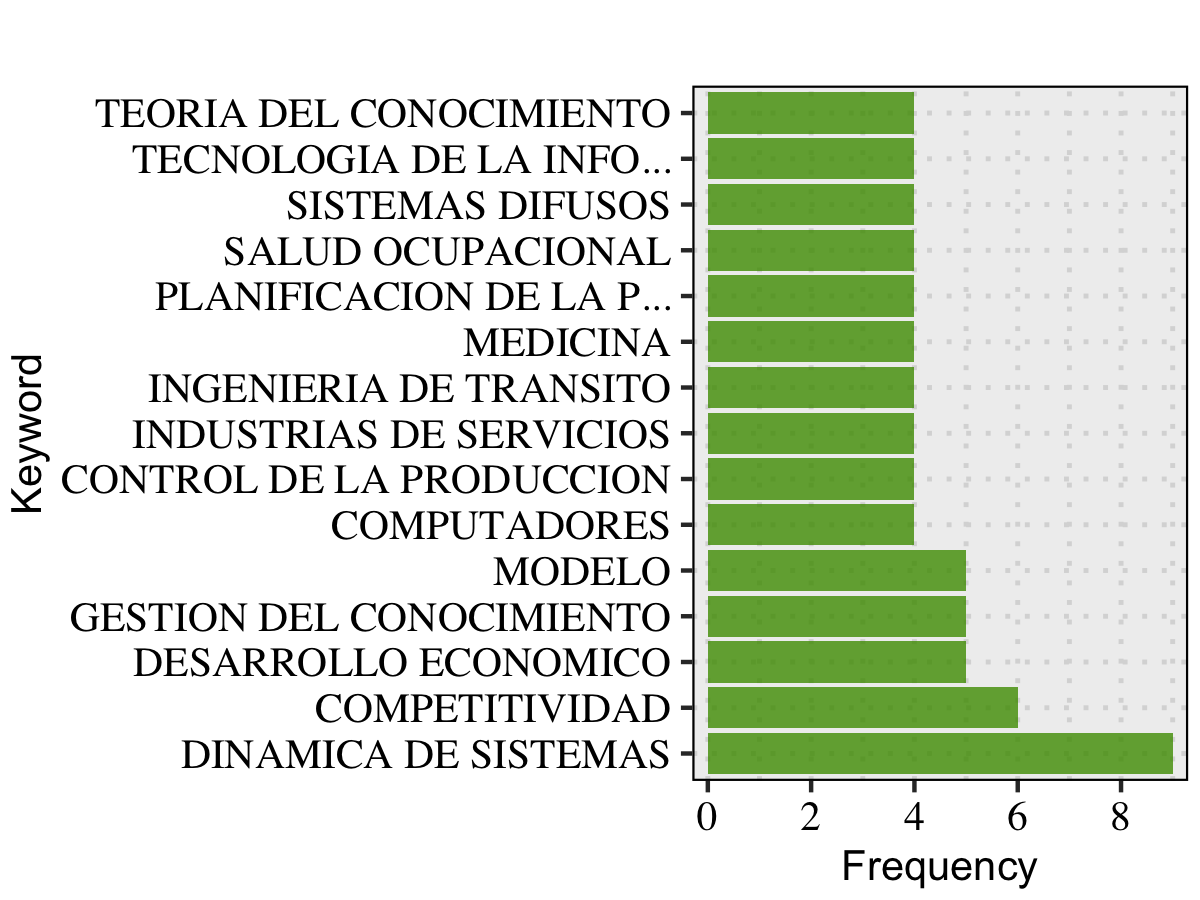}
        \caption{Field: Keywords}
     \end{subfigure}
     \hfill
     \begin{subfigure}[b]{0.5\textwidth}
        \includegraphics[scale=.18]{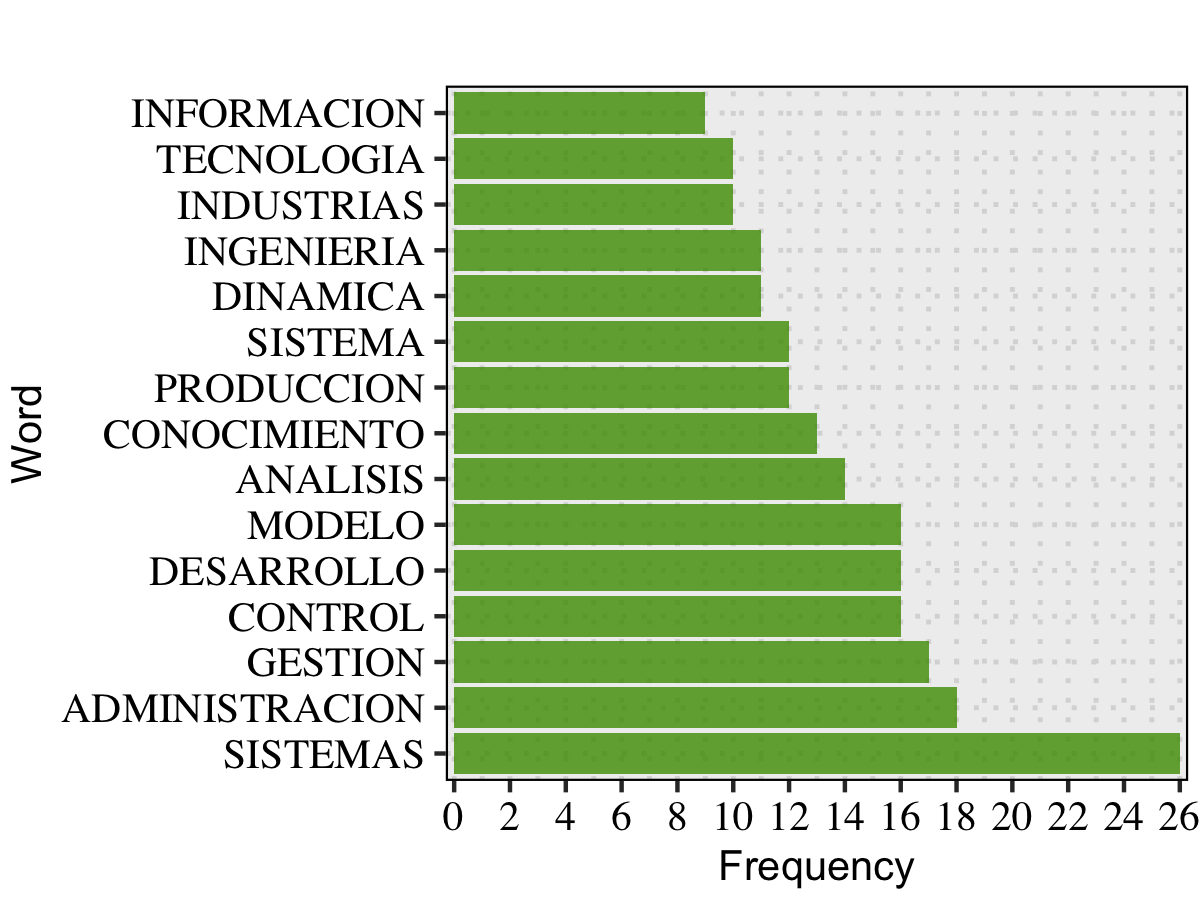}
        \caption{Field: Unigram keywords}
     \end{subfigure}

     \begin{subfigure}[b]{0.5\textwidth}
        \includegraphics[scale=.18]{IIND/frequent-words-titles.png}
        \caption{Field: Title}
     \end{subfigure}
     \hfill
     \begin{subfigure}[b]{0.5\textwidth}
        \includegraphics[scale=.18]{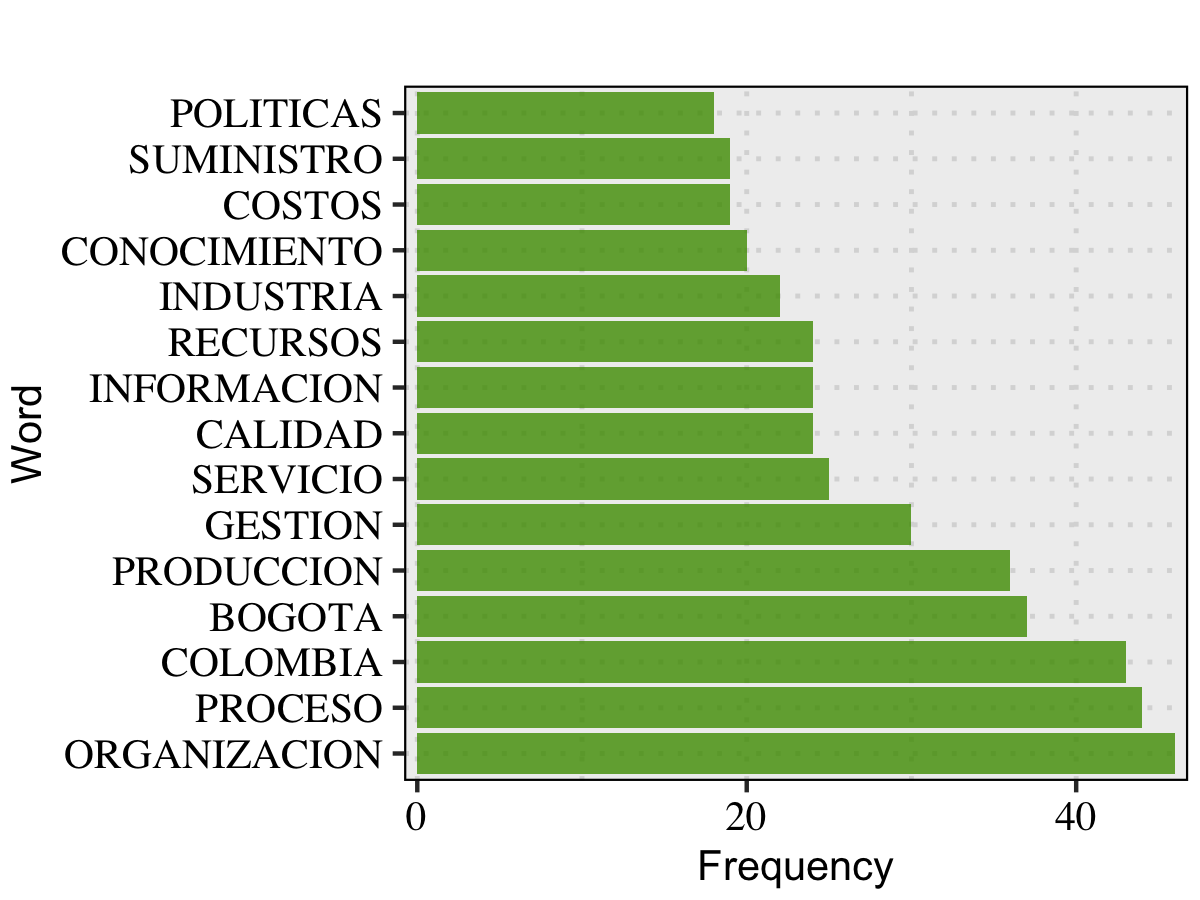}
        \caption{Field: Abstract}
     \end{subfigure}

        \caption{Frequent words analysis using different fields for the MIE dataset.}
        \label{fig:frequent-words}
\end{figure}

\subsection{Wordclouds}
\begin{figure}[H]
     \hspace{-1.5cm}
     \begin{subfigure}[b]{0.5\textwidth}
        \includegraphics[scale=.3]{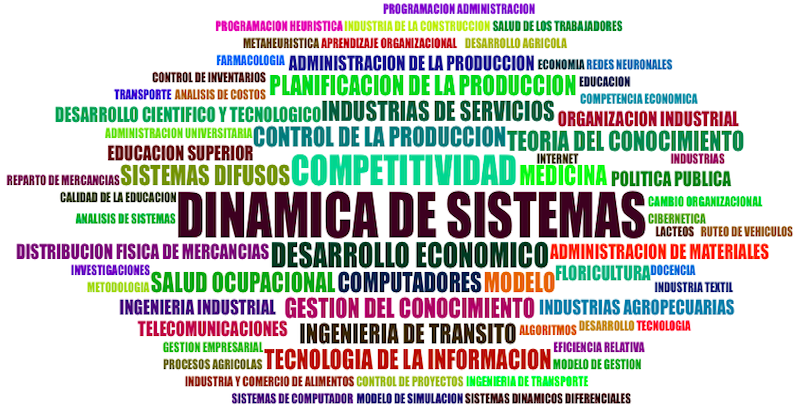}
        \caption{Field: Keywords}
     \end{subfigure}
     \hfill
     \begin{subfigure}[b]{0.5\textwidth}
        \includegraphics[scale=.3]{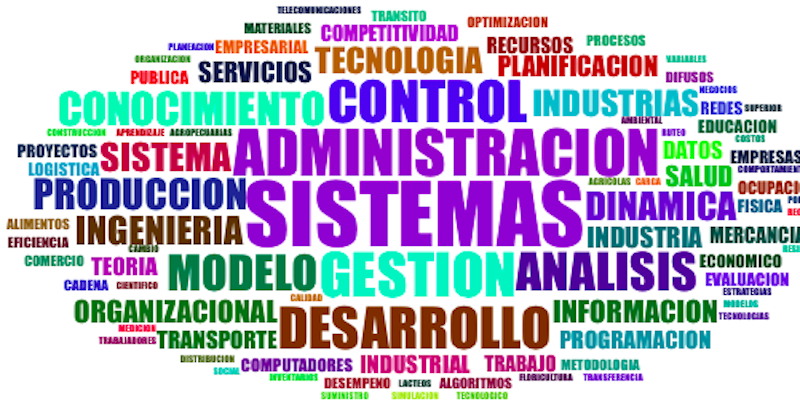}
        \caption{Field: Unigram keywords}
     \end{subfigure}

     \hspace{-1.5cm}
     \begin{subfigure}[b]{0.5\textwidth}
        \includegraphics[scale=.3]{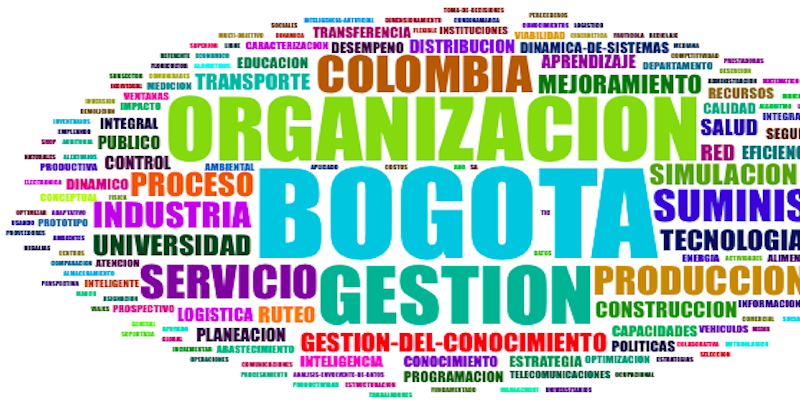}
        \caption{Field: Title}
     \end{subfigure}
     \hfill
     \begin{subfigure}[b]{0.5\textwidth}
        \includegraphics[scale=.3]{IIND/wordcloud-abstracts.png}
        \caption{Field: Abstract}
     \end{subfigure}

        \caption{Wordcloud analysis using different fields for the MIE dataset.}
        \label{fig:wordclouds}
\end{figure}

\subsection{Topic maps}
\begin{figure}[H]
     \begin{subfigure}[b]{0.5\textwidth}
        \includegraphics[scale=.18]{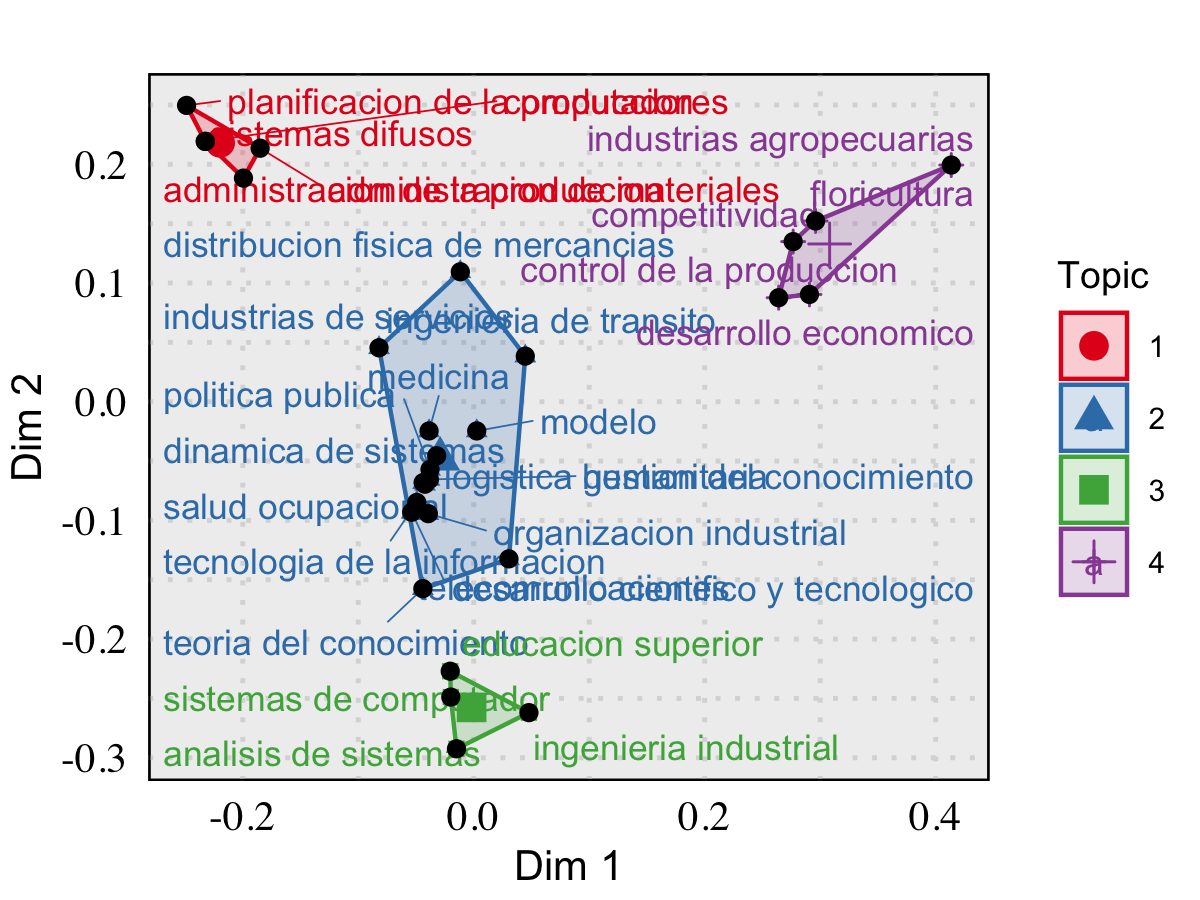}
        \caption{Field: Keywords}
     \end{subfigure}
     \hfill
     \begin{subfigure}[b]{0.5\textwidth}
        \includegraphics[scale=.18]{IIND/topic-map-id-keywords.png}
        \caption{Field: Unigram keywords}
     \end{subfigure}

     \begin{subfigure}[b]{0.5\textwidth}
        \includegraphics[scale=.18]{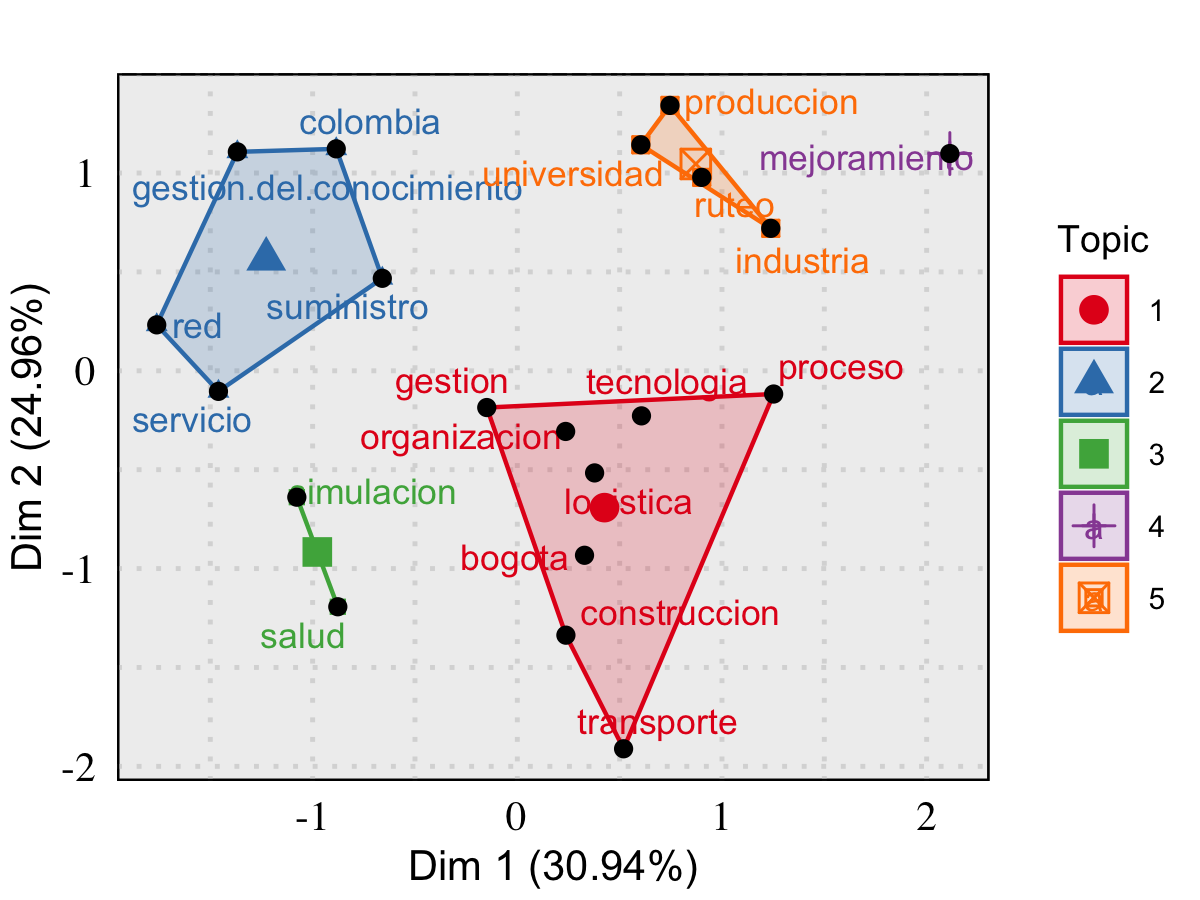}
        \caption{Field: Title}
     \end{subfigure}
     \hfill
     \begin{subfigure}[b]{0.5\textwidth}
        \includegraphics[scale=.18]{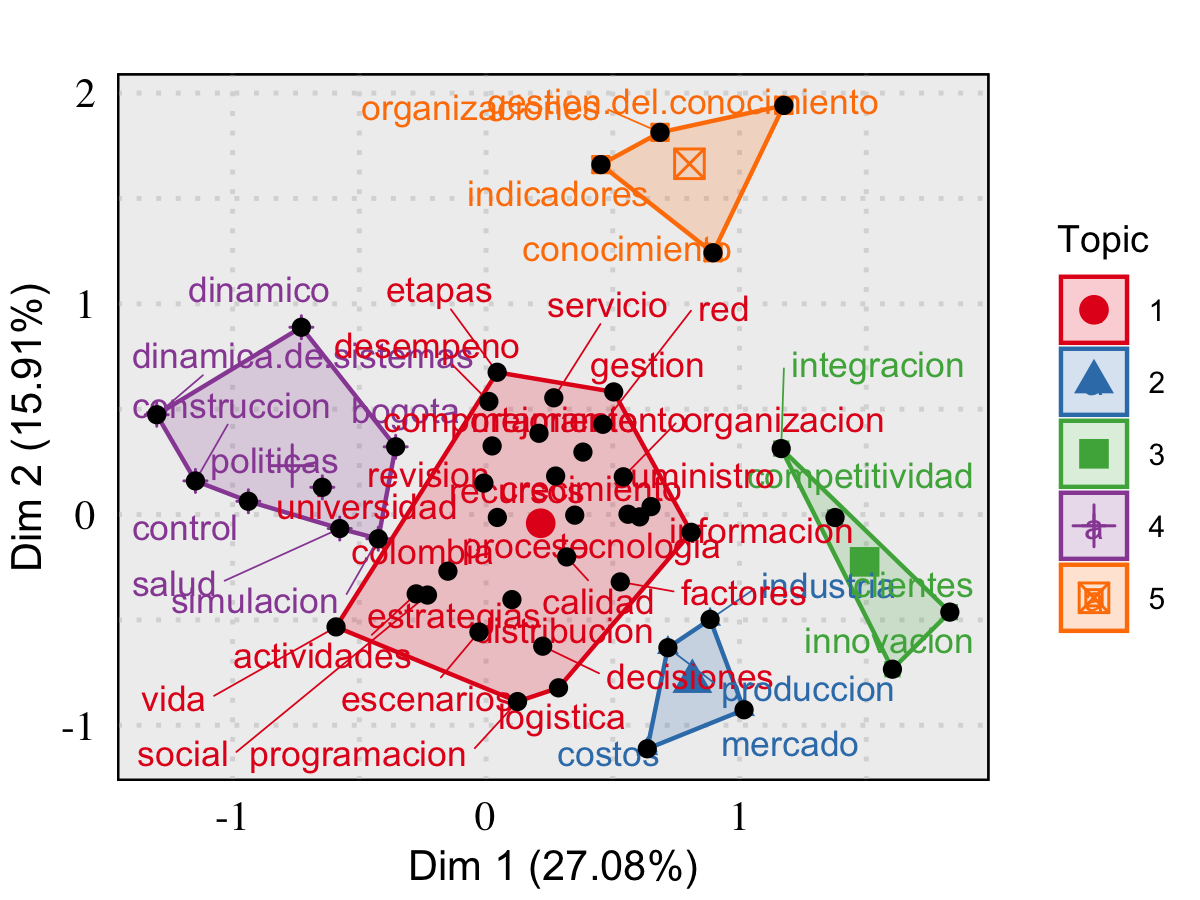}
        \caption{Field: Abstract}
     \end{subfigure}

        \caption{Topic maps generated from  different fields for the MIE dataset.}
        \label{fig:topic-maps}
\end{figure}

\subsection{Dendrograms}
\begin{figure}[H]
     \begin{subfigure}[b]{0.5\textwidth}
        \includegraphics[scale=.18]{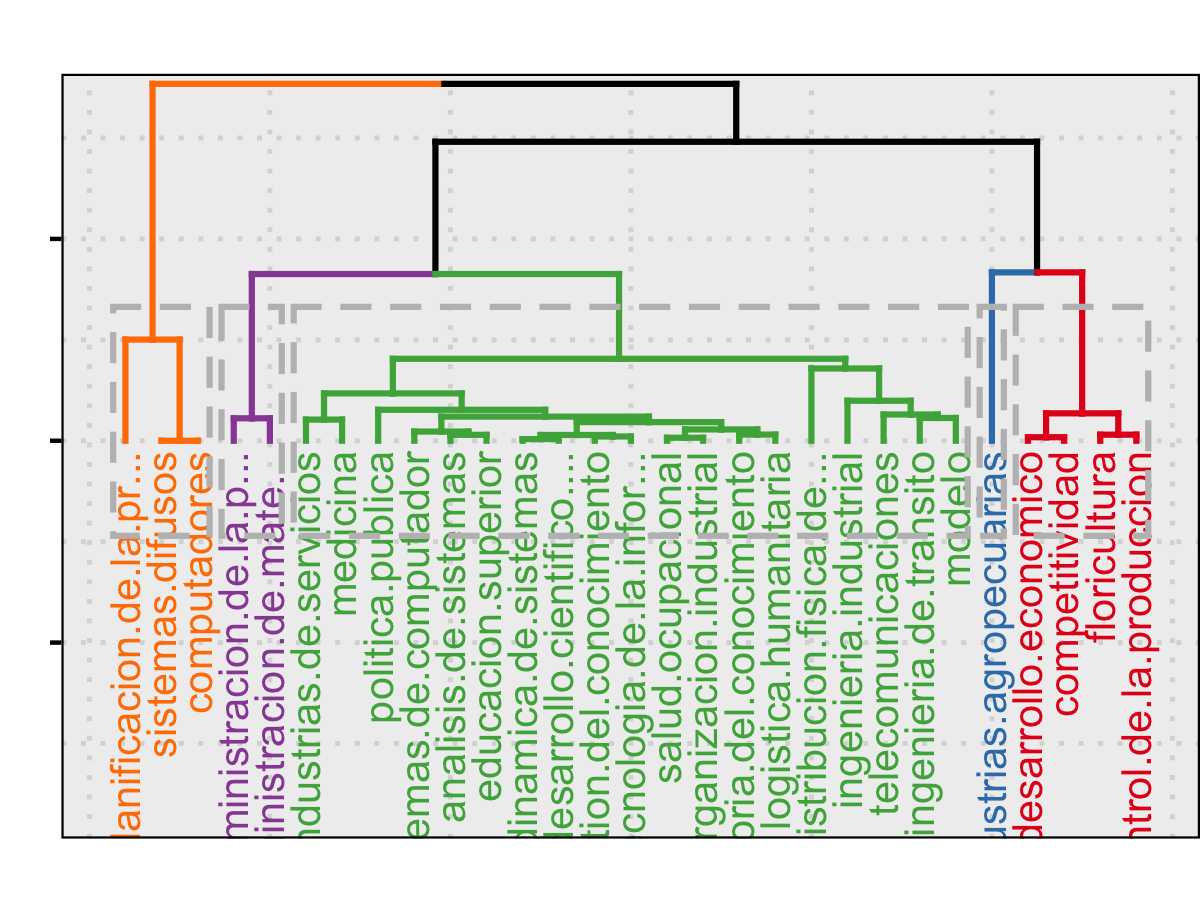}
        \caption{Field: Keywords}
     \end{subfigure}
     \hfill
     \begin{subfigure}[b]{0.5\textwidth}
        \includegraphics[scale=.18]{IIND/dendrogram-id-keywords.png}
        \caption{Field: Unigram keywords}
     \end{subfigure}

     \begin{subfigure}[b]{0.5\textwidth}
        \includegraphics[scale=.18]{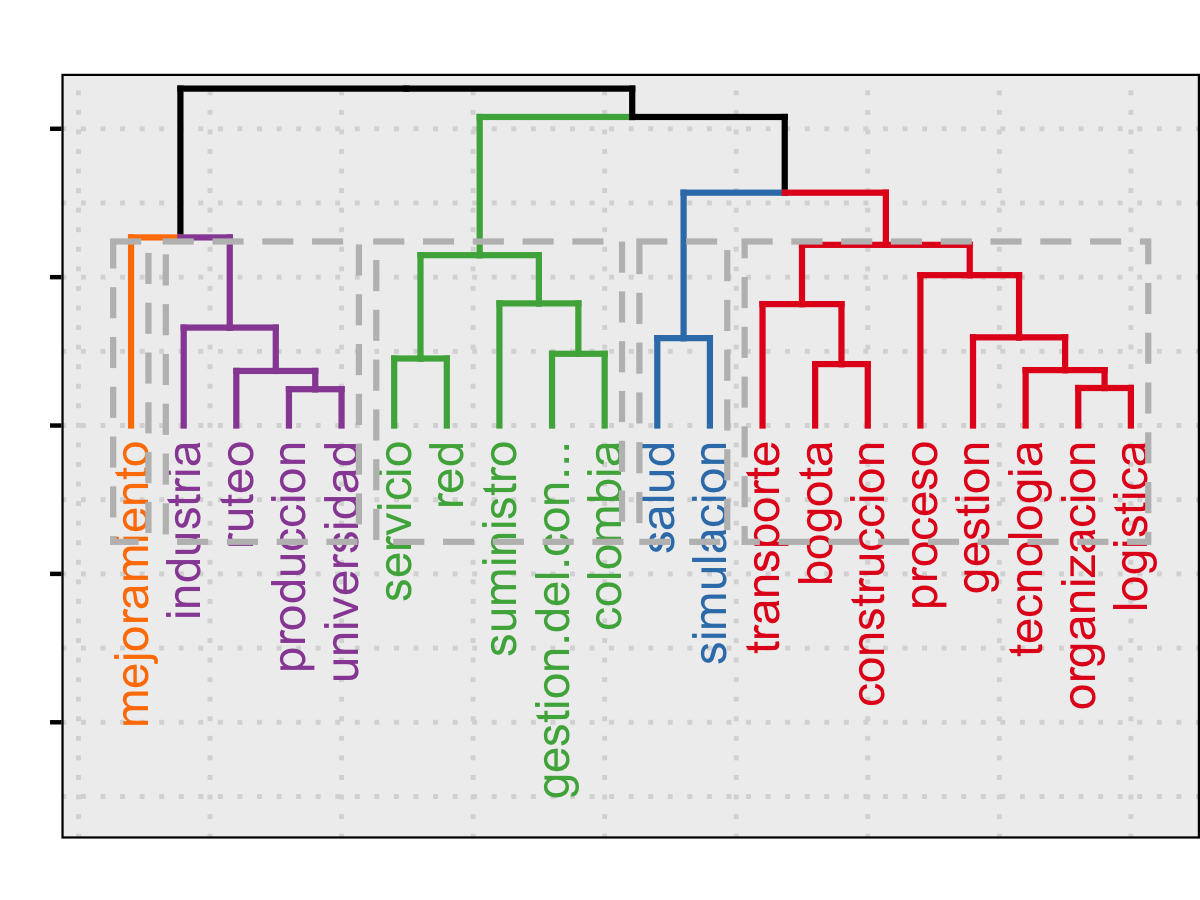}
        \caption{Field: Title}
     \end{subfigure}
     \hfill
     \begin{subfigure}[b]{0.5\textwidth}
        \includegraphics[scale=.18]{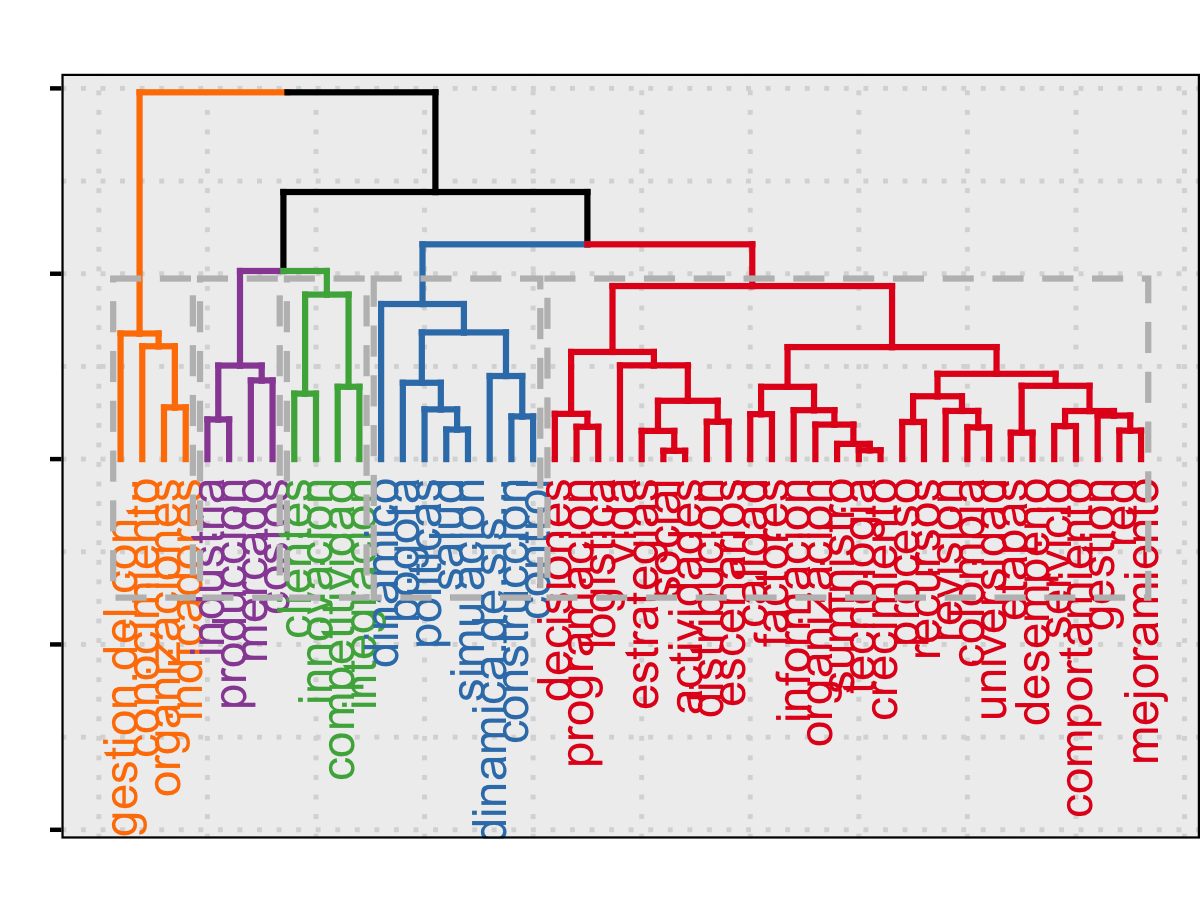}
        \caption{Field: Abstract}
     \end{subfigure}

        \caption{Dendrograms obtained from  different fields for the MIE dataset.}
        \label{fig:dendrograms}
\end{figure}

\subsection{Co-occurrence networks}
\begin{figure}[H]
     \begin{subfigure}[b]{0.5\textwidth}
        \includegraphics[scale=.4]{IIND/coocurrence-keywords.png}
        \caption{Field: Keywords}
     \end{subfigure}
     \hfill
     \begin{subfigure}[b]{0.5\textwidth}
        \includegraphics[scale=.4]{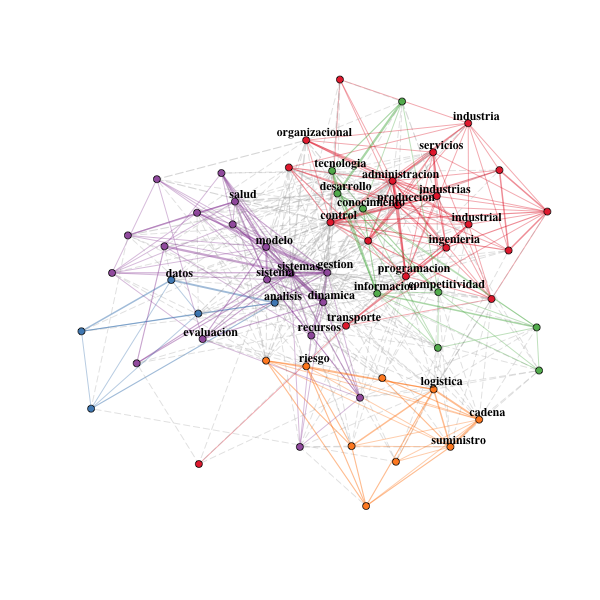}
        \caption{Field: Unigram keywords}
     \end{subfigure}

     \begin{subfigure}[b]{0.5\textwidth}
        \includegraphics[scale=.4]{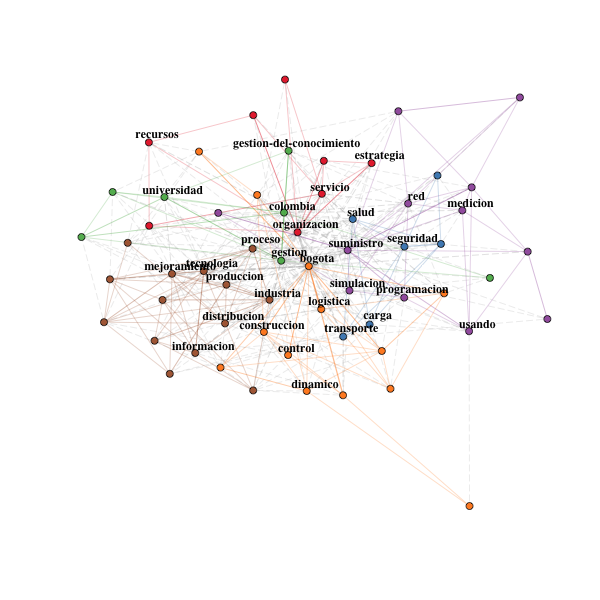}
        \caption{Field: Title}
     \end{subfigure}
     \hfill
     \begin{subfigure}[b]{0.5\textwidth}
        \includegraphics[scale=.4]{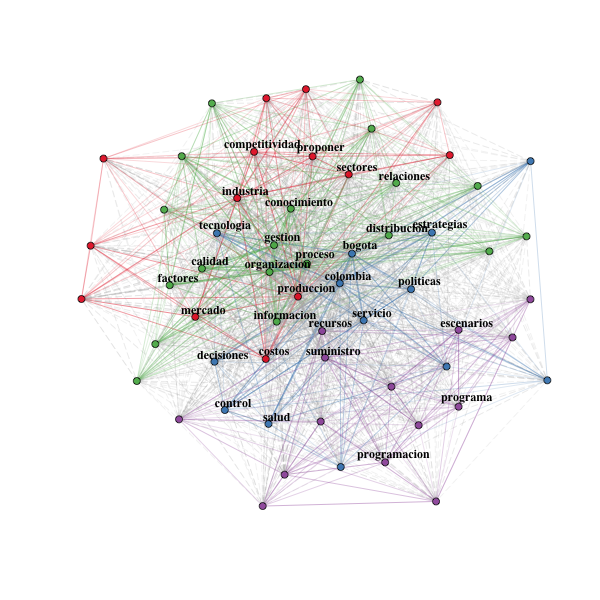}
        \caption{Field: Abstract}
     \end{subfigure}

        \caption{Co-occurrence networks from  different fields for the MIE dataset.}
        \label{fig:cooccurrence}
\end{figure}

\subsection{Thematic maps}
\begin{figure}[H]
     \begin{subfigure}[b]{0.5\textwidth}
        \includegraphics[scale=.12]{IIND/thematic-map-keywords.png}
        \caption{Field: Keywords}
     \end{subfigure}
     \hfill
     \begin{subfigure}[b]{0.5\textwidth}
        \includegraphics[scale=.12]{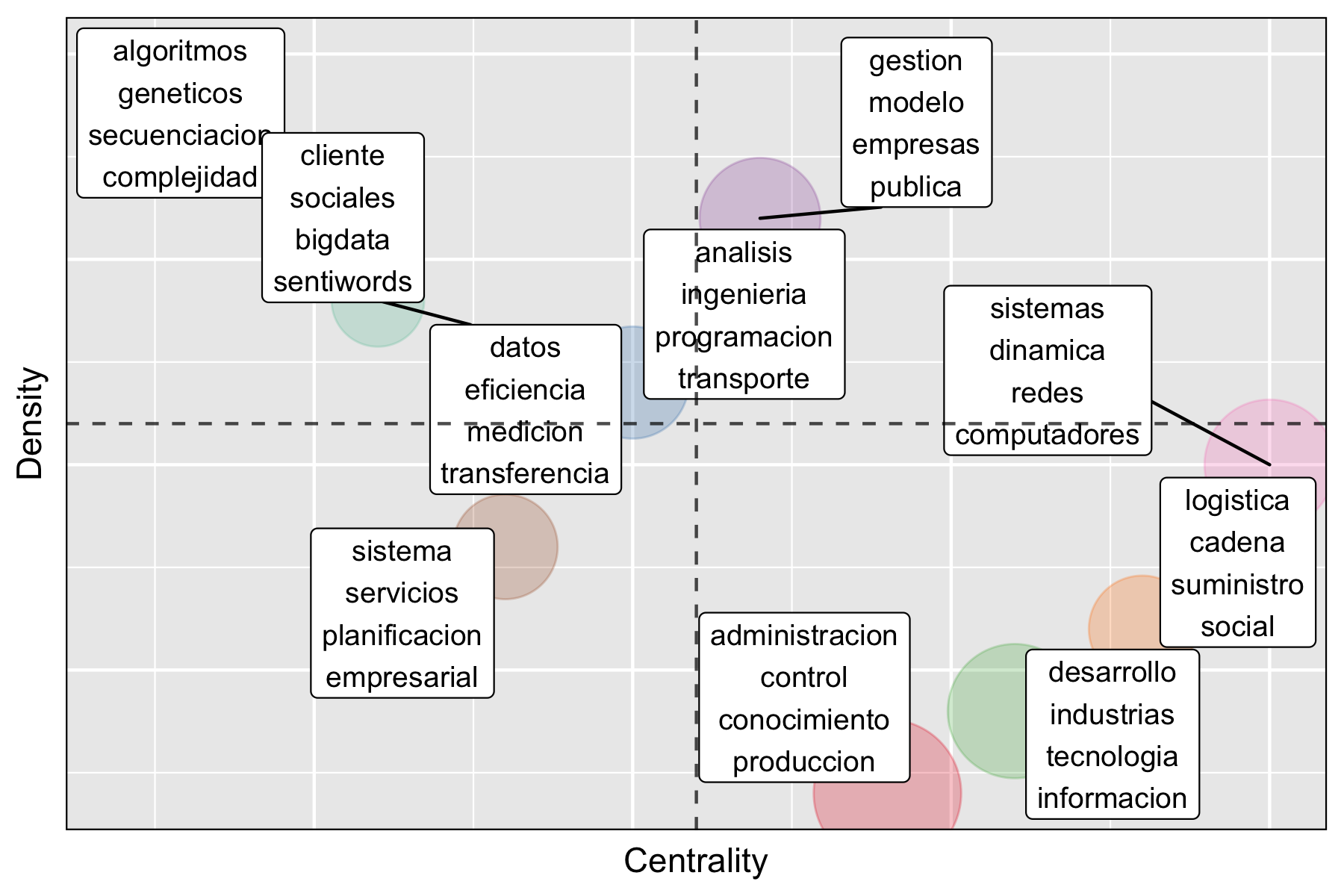}
        \caption{Field: Unigram keywords}
     \end{subfigure}

        \caption{Thematic maps generated from  different fields for the MIE dataset.}
        \label{fig:thematic-maps}
\end{figure}

\clearpage
\section{Research landscape of the MSc. in Information Sciences, UDFJC.}
This programme was established on 1989 with a focus on the areas of Telecommunications and Information Systems. On top of these two majors, since 2011 the programme has been reformed to open up new lines of research, including Geomatics, Software Engineering and Artificial Intelligence. 
The bibliographic corpus used to conduct the analysis, consisted of the metadata of dissertations completed during the period 2012-2020, gathered according to the guidelines provided in \secref{sec:dataset}. We refer to this corpus as the \textit{MIS} dataset. The results of each stage in the proposed workflow are reported next.

\medskip
\noindent\textbf{\emph{RQ1. Production dynamics}}

\begin{table}[H]
    \footnotesize
    \centering
    \setlength{\tabcolsep}{.2cm}
    \def\arraystretch{1.15}

    \begin{tabular}{|l|c|l|c|}
    \hline
    \multicolumn{2}{|c|}{\textbf{Dynamics}}     & \multicolumn{2}{c|}{\textbf{Structure}} \\ \hline
    Timespan                        & 2012-2020 & Authors                       & 243     \\ \hline
    Documents                       & 170       & Author appearances            & 355     \\ \hline
    Avg. citations per document     & 0.18      & Single-authored documents     & 1       \\ \hline
    Avg. citations per year per doc & 0.03      & Authors per document          & 1.43    \\ \hline
    Author's keywords               & 656       & Co-authors per documents      & 2.09    \\ \hline
    Unigram keywords                & 669       & Collaboration Index           & 1.43    \\ \hline
    Avg. dissertations per year     & 18.9      & References*                    & 5169  \\ \hline
    \multicolumn{4}{r}{*\scriptsize{References were only available since 2016 (66 documents)}} 
    \end{tabular}
    
        \caption{Bibliometric statistics for the MIS dataset.}
        \label{tab:statistics-MIS}
\end{table}

\begin{figure}[H]
     \centering
     \begin{subfigure}[b]{0.45\textwidth}
         \centering
         \includegraphics[scale=.15]{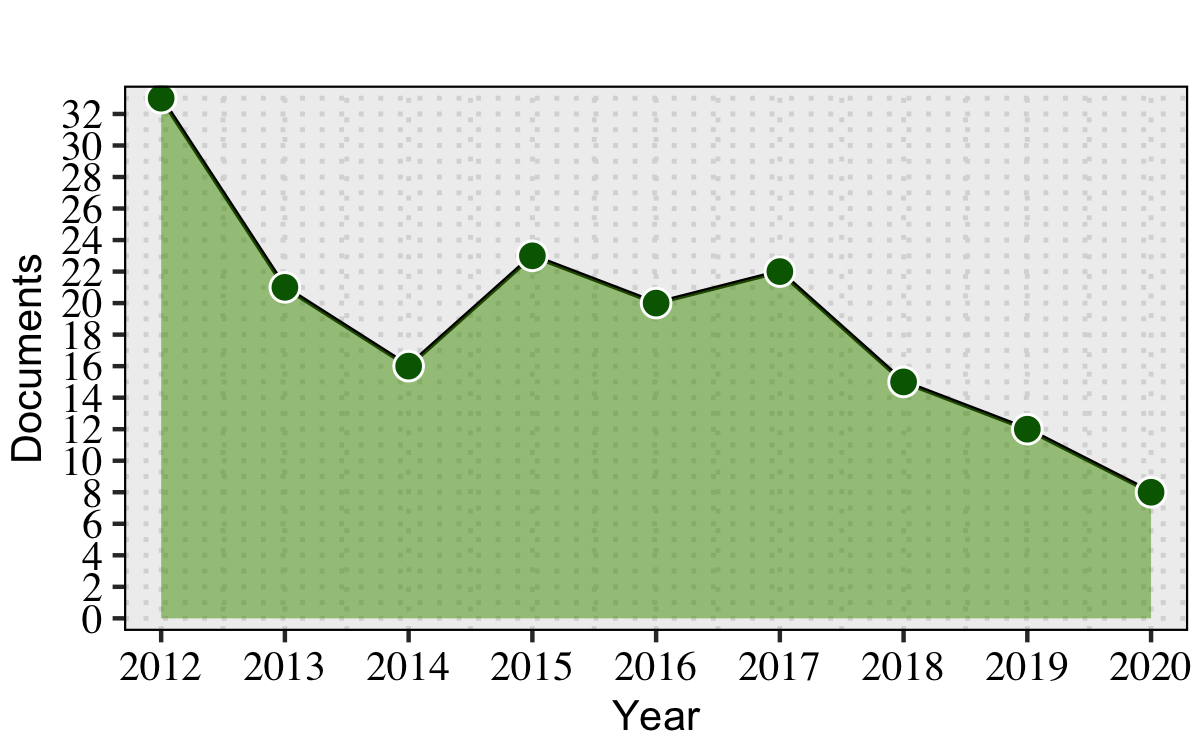}
         \caption{Production growth}
         \label{cic:production-growth}
     \end{subfigure}
     \hfill
     \begin{subfigure}[b]{0.45\textwidth}
         \centering
         \includegraphics[scale=.15]{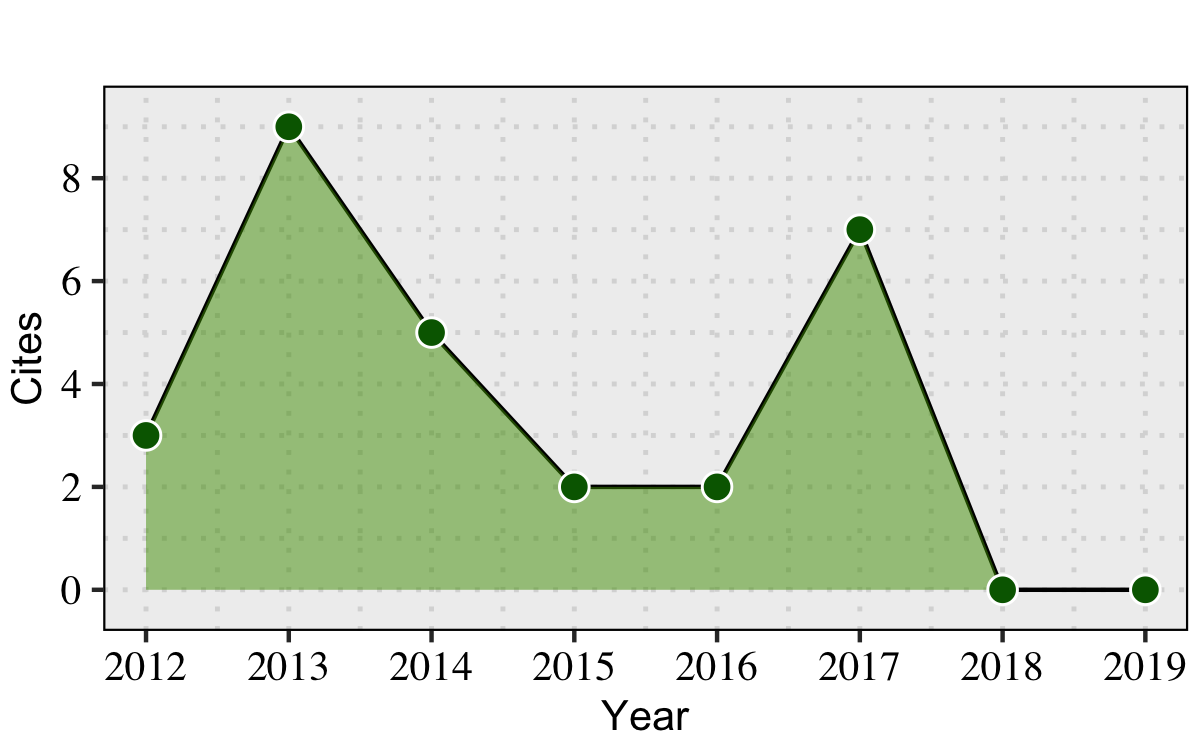}
         \caption{Citation count}
         \label{cic:citation-count}
     \end{subfigure}
     \hfill \\
     
     \begin{subfigure}[b]{0.45\textwidth}
         \centering
        \includegraphics[scale=.15]{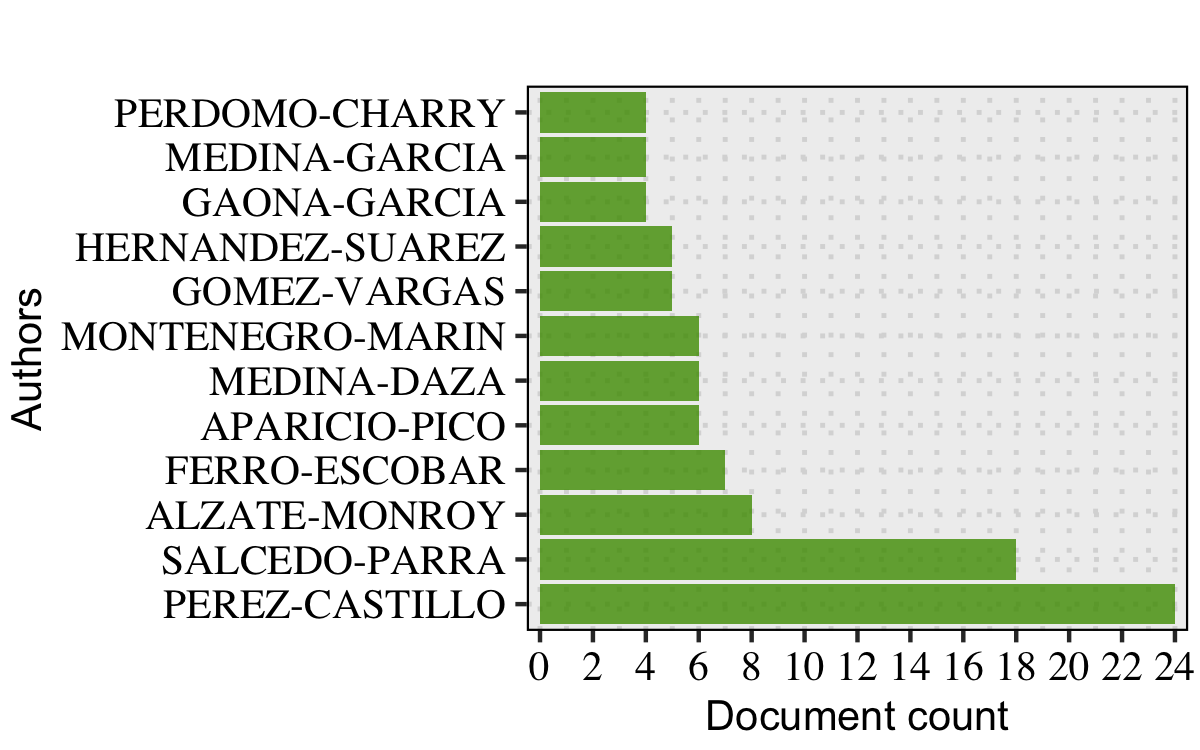}
        \caption{Production distribution (authors)}
        \label{cic:production-distribution-authors}
     \end{subfigure}
     \hfill
     \begin{subfigure}[b]{0.45\textwidth}
         \centering
         \includegraphics[scale=.15]{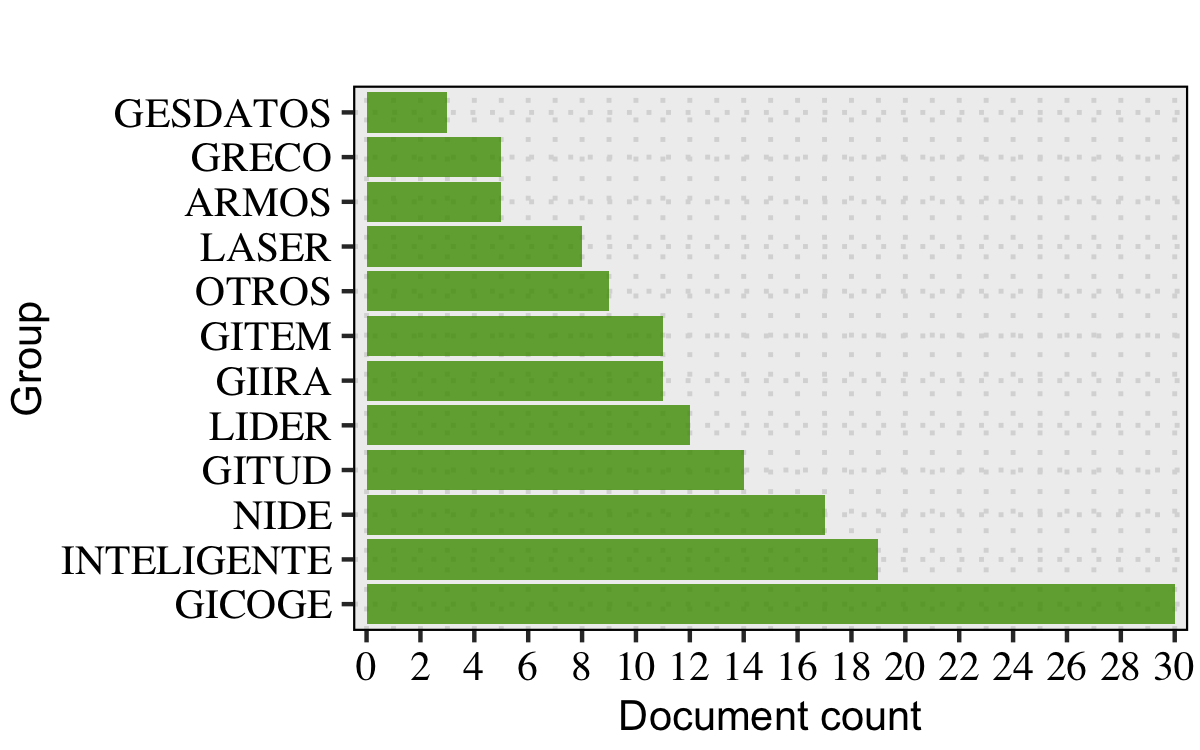}
         \caption{Production distribution (groups)}
         \label{cic:production-distribution-groups}
     \end{subfigure}
     \hfill \\
          
     \begin{subfigure}[b]{0.45\textwidth}
         \centering
        \includegraphics[scale=.15]{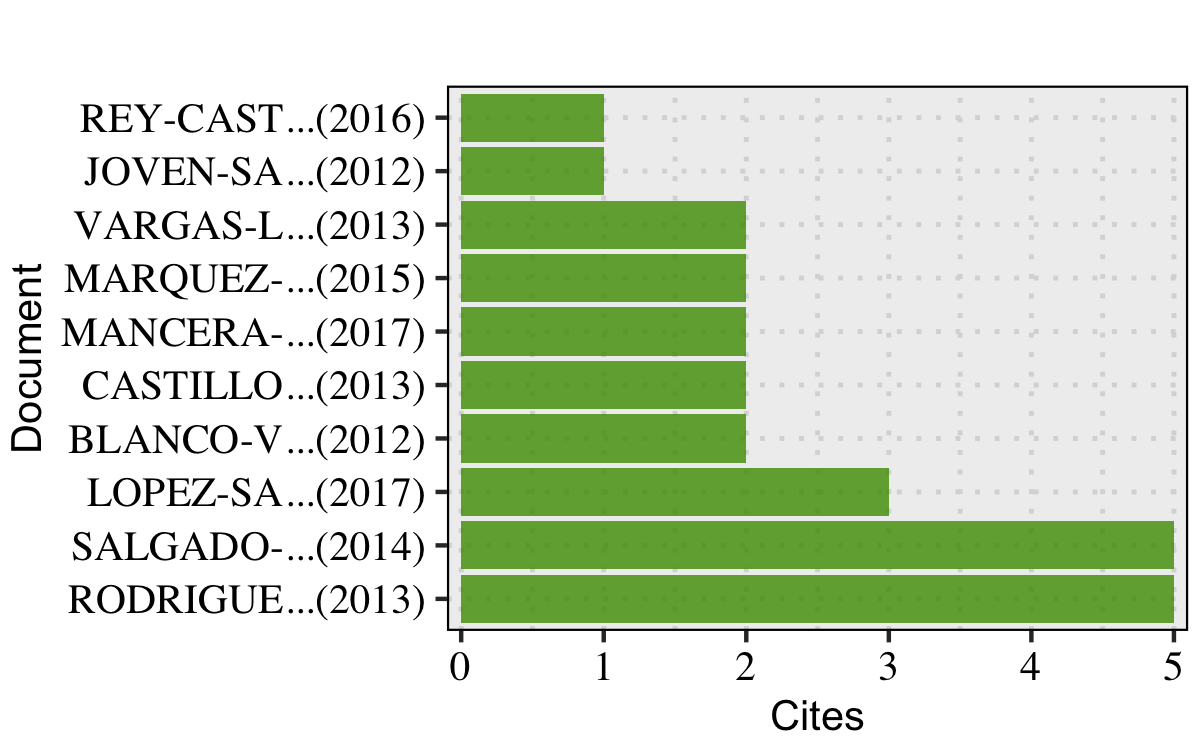}
        \caption{Citation distribution}
        \label{cic:citation-distribution}
     \end{subfigure}
     \hfill 
     \begin{subfigure}[b]{0.45\textwidth}
         \centering
         \includegraphics[scale=.15]{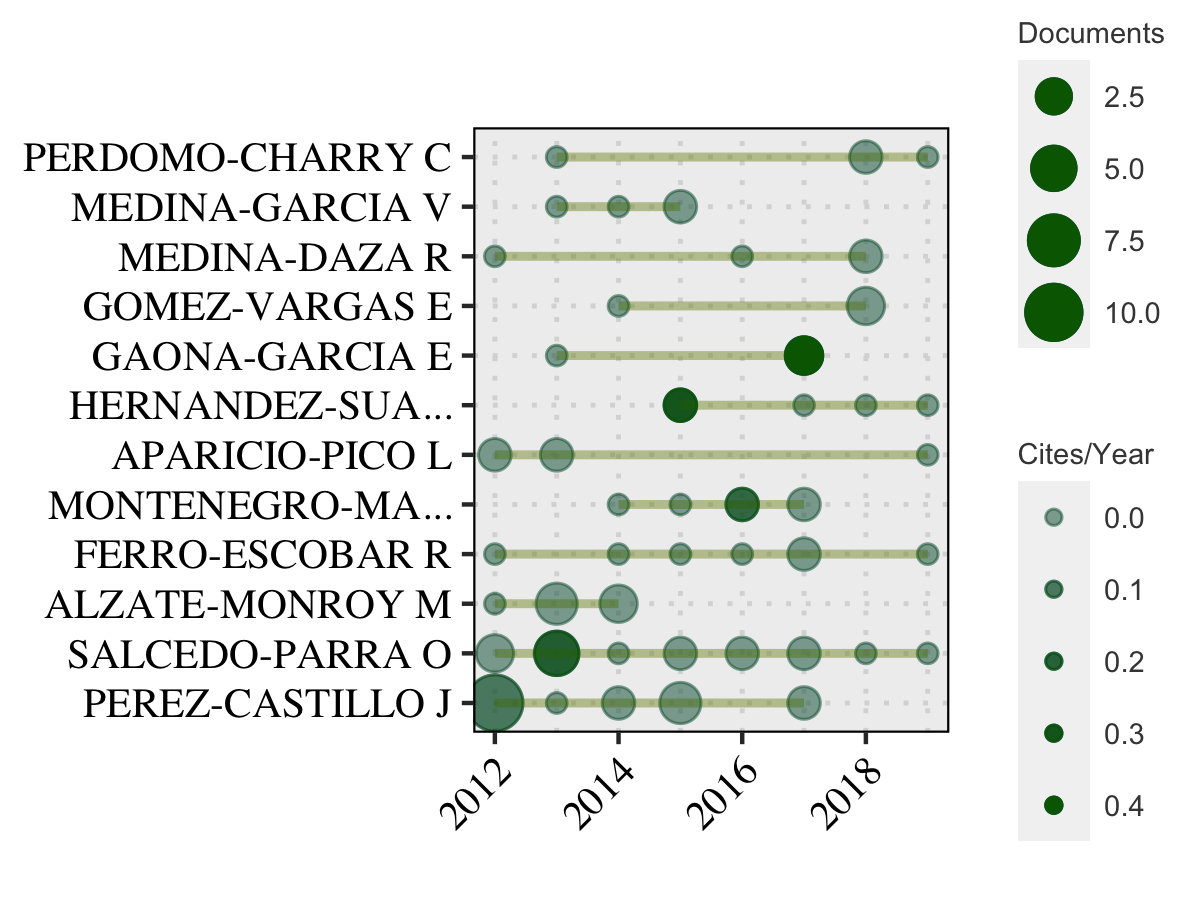}
         \caption{Author's timelines}
         \label{cic:authors-timelines}
     \end{subfigure}
     \hfill \\
     
     \begin{subfigure}[b]{0.45\textwidth}
         \centering
         \includegraphics[scale=.15]{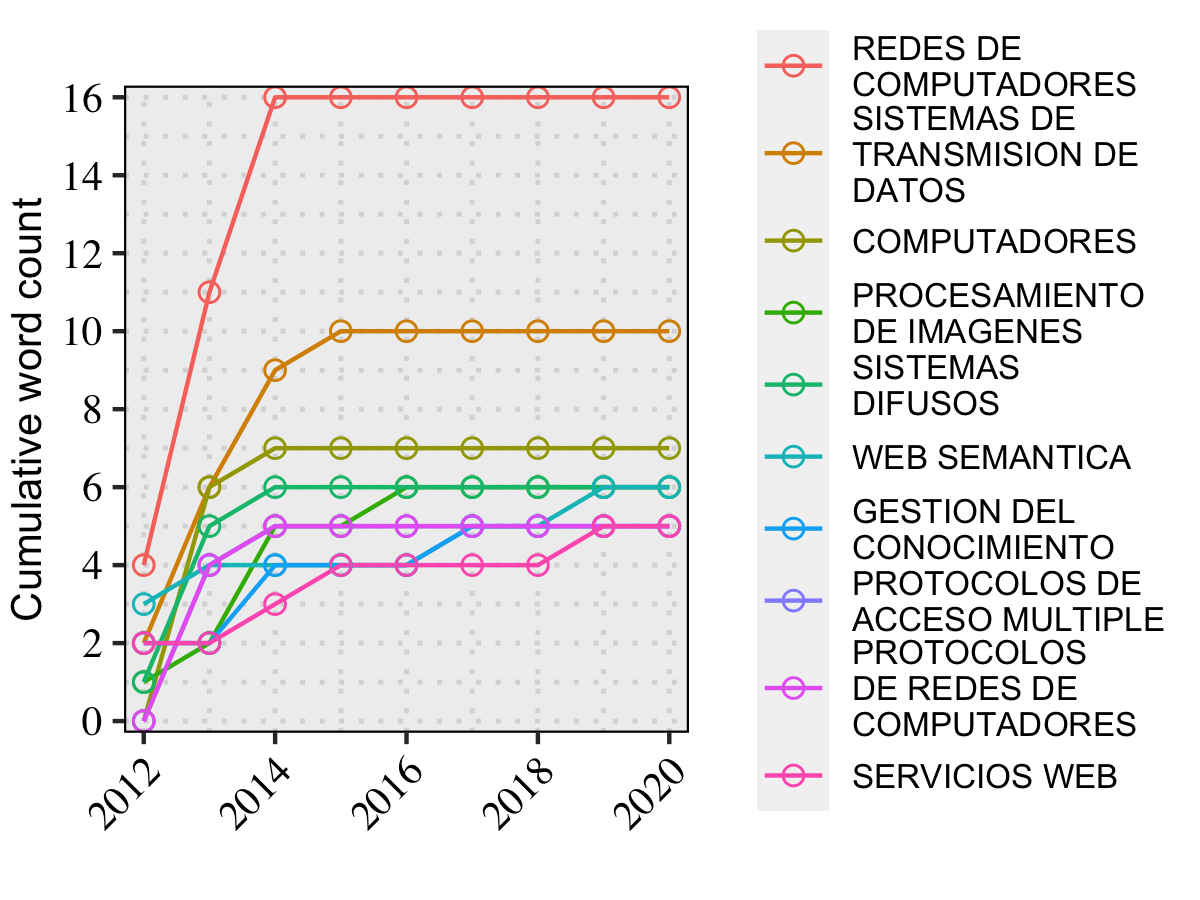}
         \caption{Word trends (keywords)}
         \label{cic:word-trends-keywords}
     \end{subfigure}
     \hfill
     \begin{subfigure}[b]{0.45\textwidth}
         \centering
         \includegraphics[scale=.15]{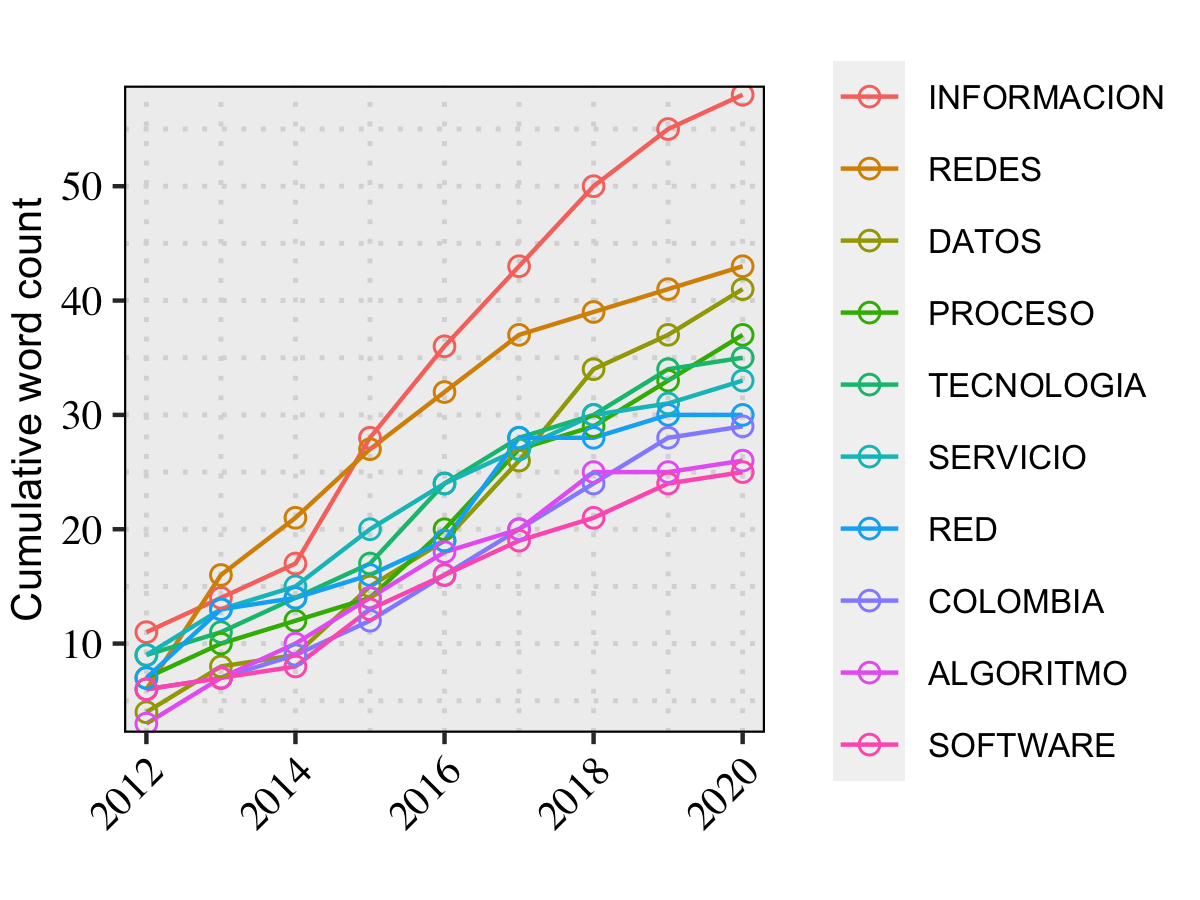}
         \caption{Word trends (abstracts)}
         \label{cic:word-trends-abstract}
     \end{subfigure}
     \hfill \\     
     
        \caption{Results of the RQ1 analysis (production dynamics) for the MIS dataset.}
        \label{fig:RQ1-MIS}
\end{figure}

\noindent\textbf{\emph{RQ2. Conceptual structures}}
\vspace{-.5cm}
\begin{figure}[H]
     \centering

     \begin{subfigure}[b]{0.45\textwidth}
        \hspace{-1cm}
        \includegraphics[scale=.15]{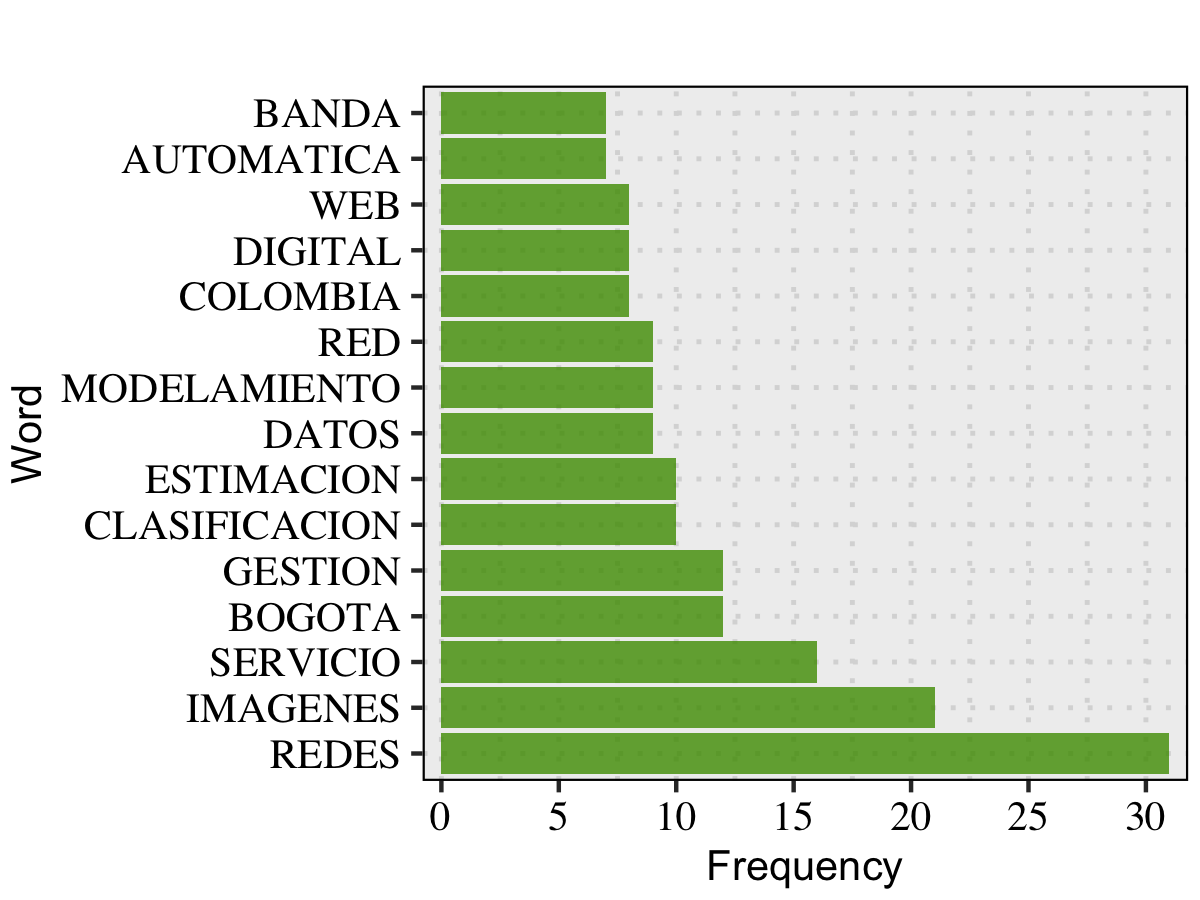}
        \caption{Frequent words (titles)}
        \label{cic:frequent-words}
     \end{subfigure}
     \hfill
     \begin{subfigure}[b]{0.45\textwidth}
         \centering
        \includegraphics[scale=.3]{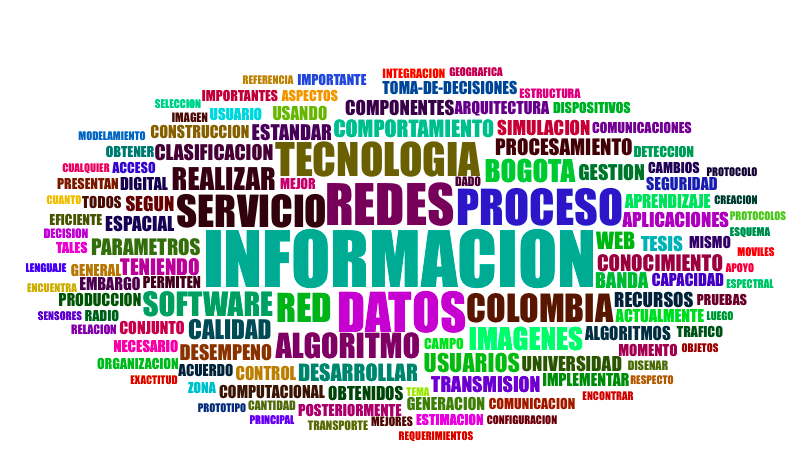}
        \caption{Word cloud (abstracts)}
        \label{cic:word-cloud}
     \end{subfigure}
     \hfill \\
     
     \begin{subfigure}[b]{0.45\textwidth}
         \centering
        \includegraphics[scale=.15]{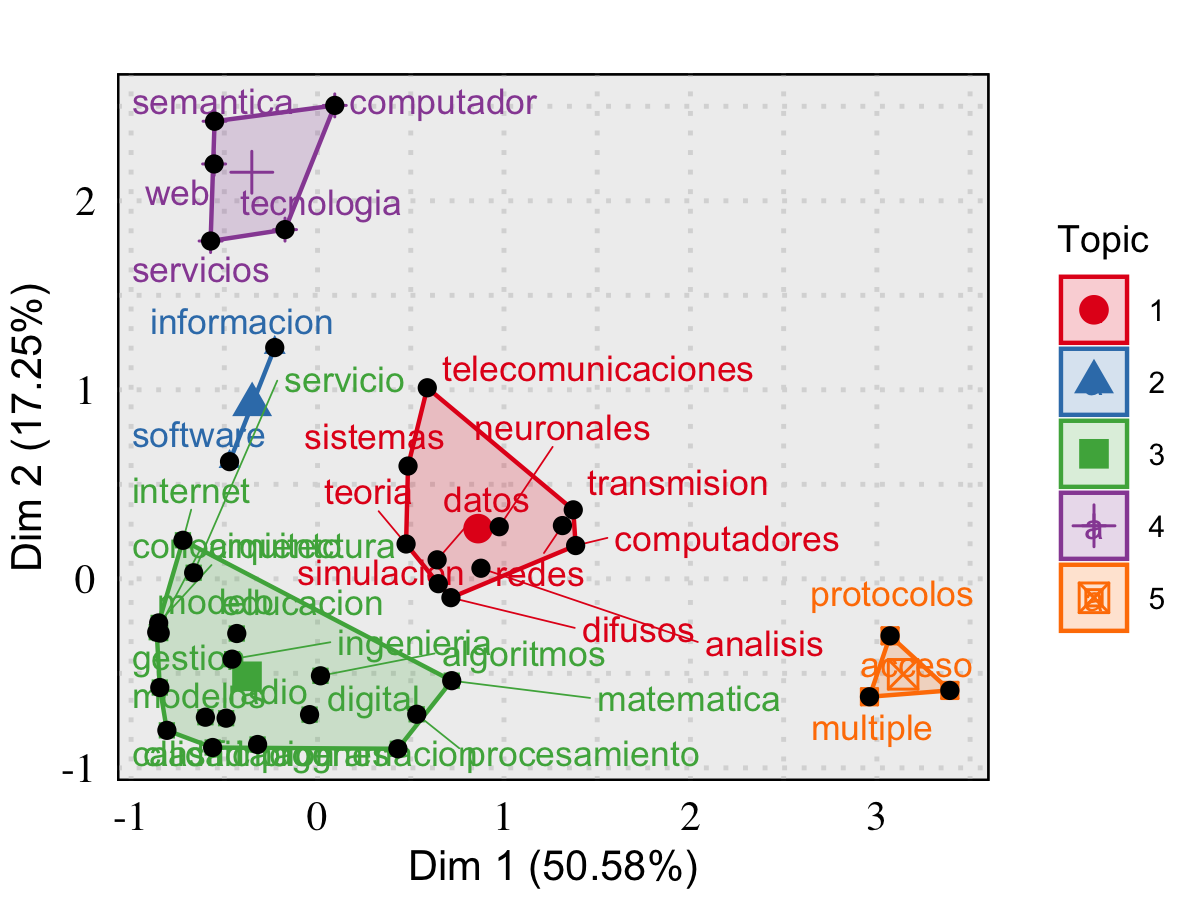}
        \caption{Topic map (keywords)}
        \label{cic:topic-map}
     \end{subfigure}
     \hfill
     \begin{subfigure}[b]{0.45\textwidth}
         \centering
        \includegraphics[scale=.15]{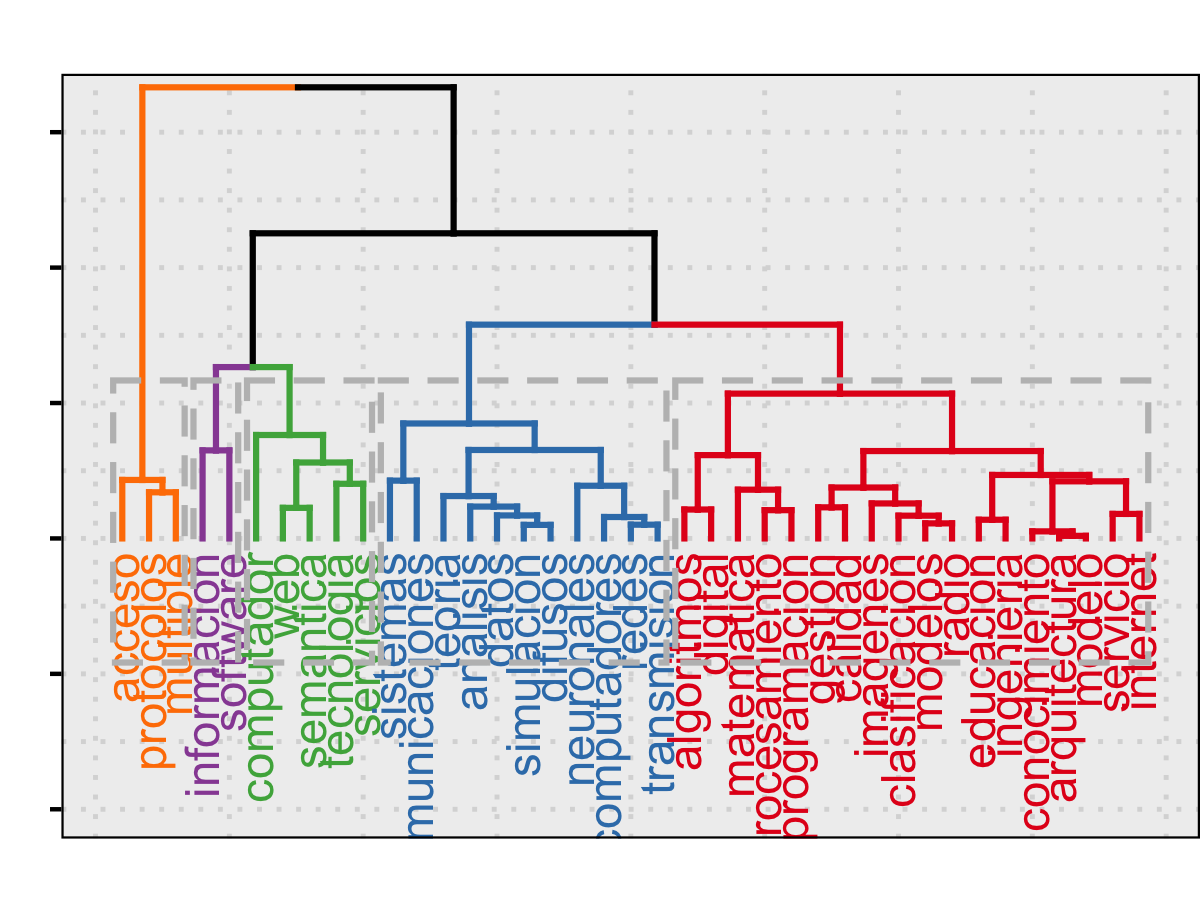}
        \caption{Word dendrogram (keywords)}
        \label{cic:word-dendrogram}
     \end{subfigure}
     \hfill \\
     
     \begin{subfigure}[b]{0.45\textwidth}
        \hspace{-2cm}
        \includegraphics[scale=.4]{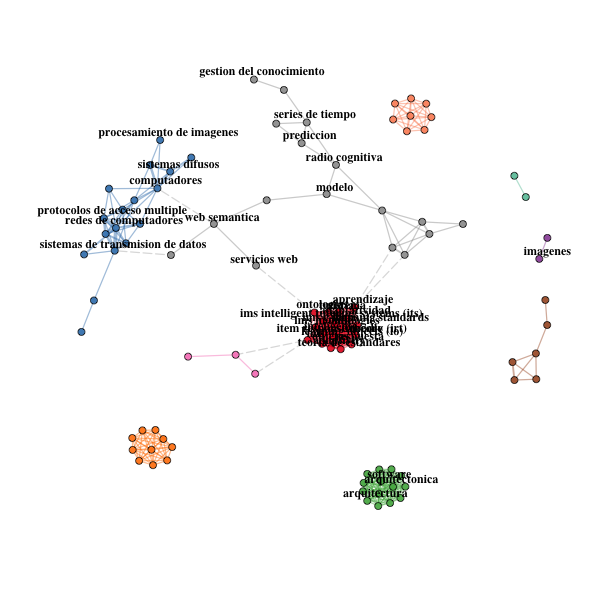}
        \caption{Co-occurrence network (keywords)}
        \label{cic:coocurrence-network}
     \end{subfigure}
     \hfill
     \begin{subfigure}[b]{0.45\textwidth}
         \centering
        \includegraphics[scale=.12]{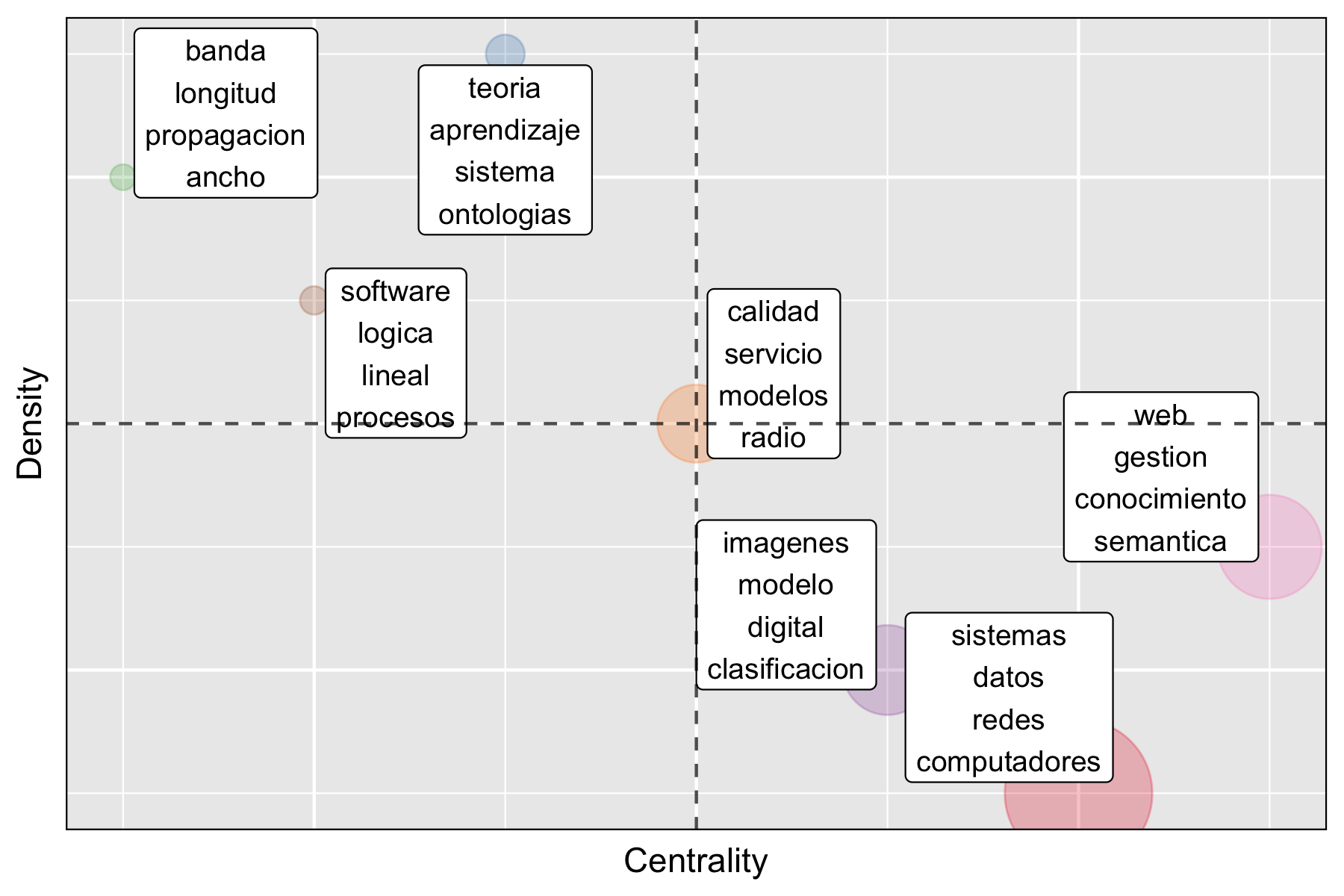}
        \caption{Thematic map (keywords)}
        \label{cic:thematic-map}
     \end{subfigure}
     
        \caption{Results of the RQ2 analysis (conceptual structures) for the MIS dataset.}
        \label{fig:RQ2-MIS}
\end{figure}

\begin{figure}[H]
    \centering
    \includegraphics[scale=.3]{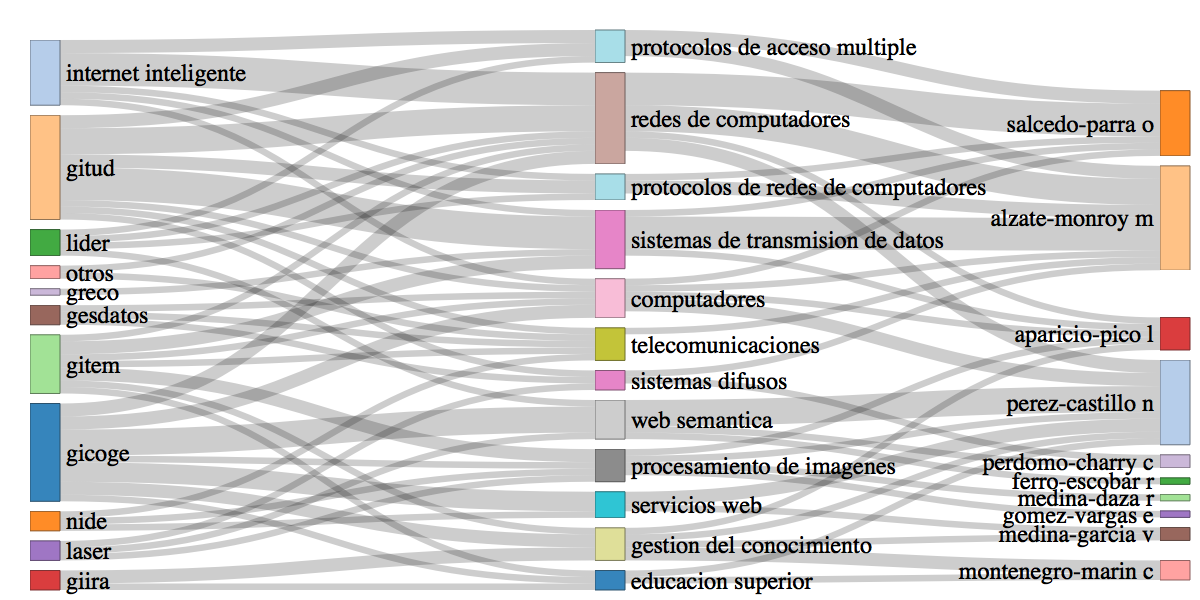}
    \caption{Energy flow through conceptual structures (MIS dataset).}
    \label{cic:energy-flow-conceptual}
\end{figure}

\noindent\textbf{\emph{RQ3. Collaboration structures}}

\begin{figure}[t]
     \centering
     \begin{subfigure}[b]{0.45\textwidth}
        \hspace{-1.5cm}
        \includegraphics[scale=.3]{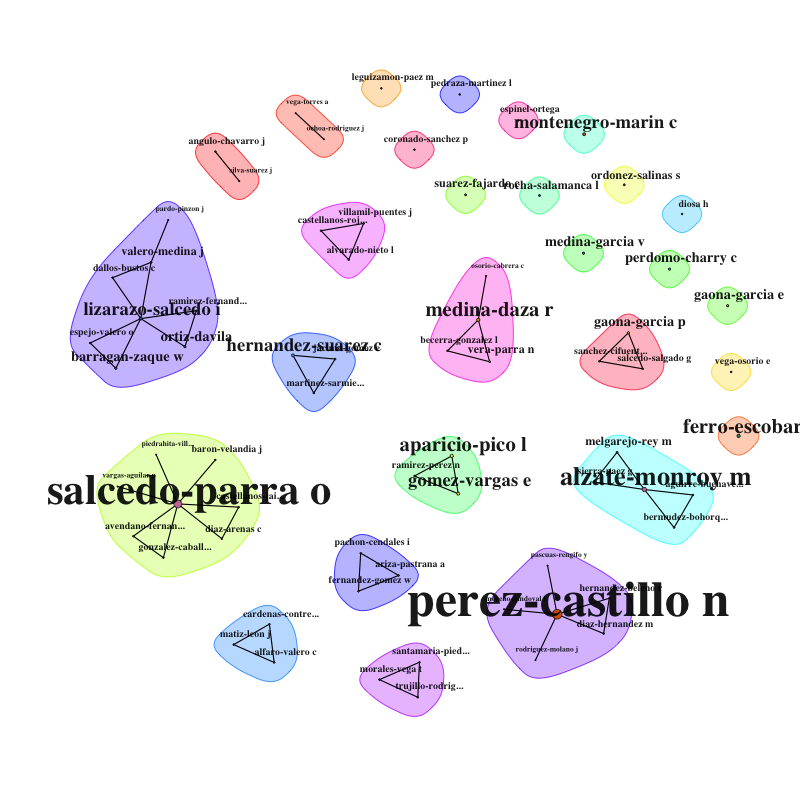}
        \caption{Authors collaboration network}
        \label{cic:collaboration-authors}
     \end{subfigure}
     \hfill
     \begin{subfigure}[b]{0.45\textwidth}
        \includegraphics[scale=.3]{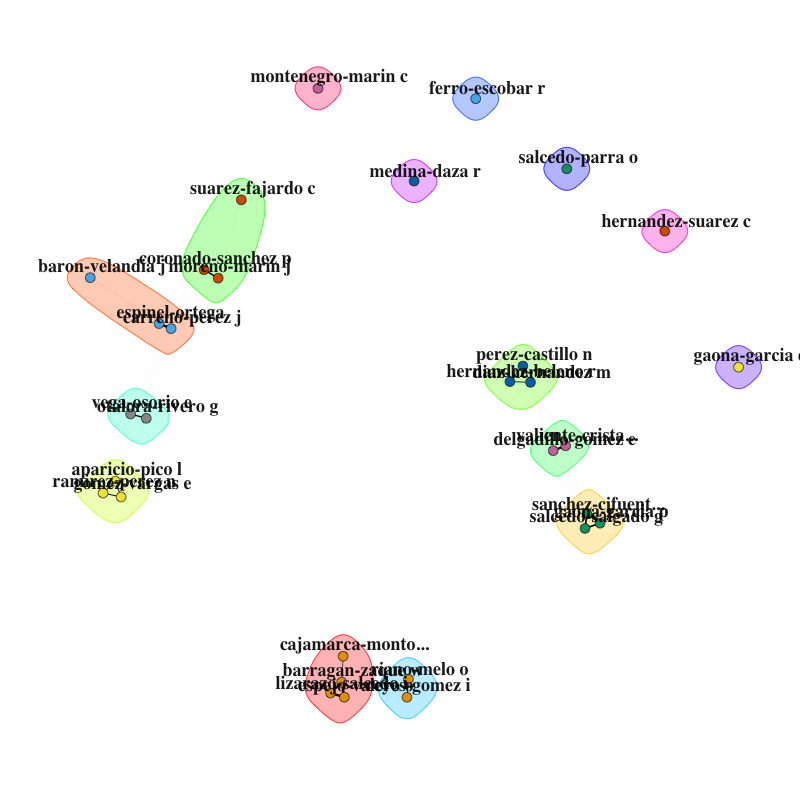}
        \caption{Authors coupling network}
        \label{cic:coupling-authors}
     \end{subfigure}

     \centering
     \begin{subfigure}[b]{0.45\textwidth}
        \hspace{-1.5cm}
        \includegraphics[scale=.3]{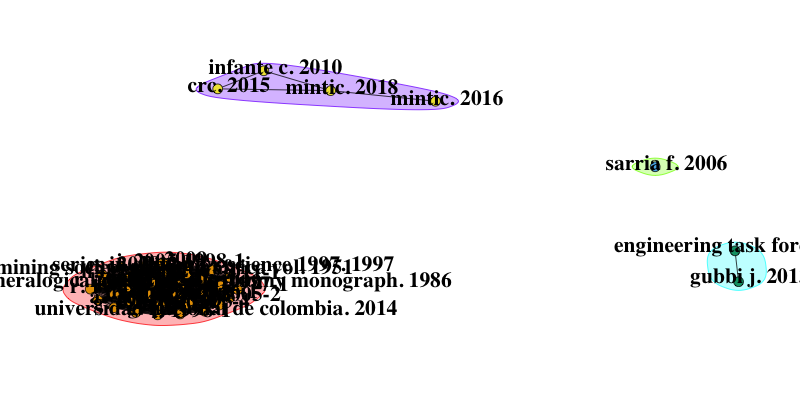}
        \caption{Co-citation references network}
        \label{cic:cocitation-references}
     \end{subfigure}
     \hfill
     \begin{subfigure}[b]{0.45\textwidth}
        \includegraphics[scale=.3]{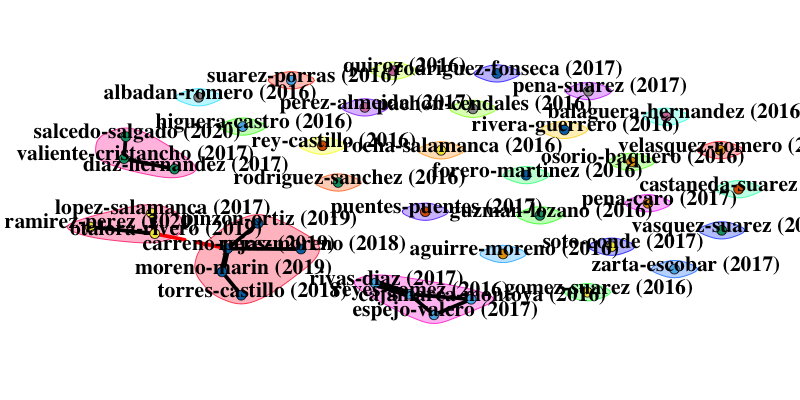}
        \caption{Manuscript coupling network}
        \label{cic:coupling-manuscripts}
     \end{subfigure}

        \caption{Results of the RQ3 analysis (collaboration structures) for the MIS dataset.}
        \label{fig:RQ3-MIS}
\end{figure}

\begin{figure}[H]
    \centering
    \includegraphics[scale=.3]{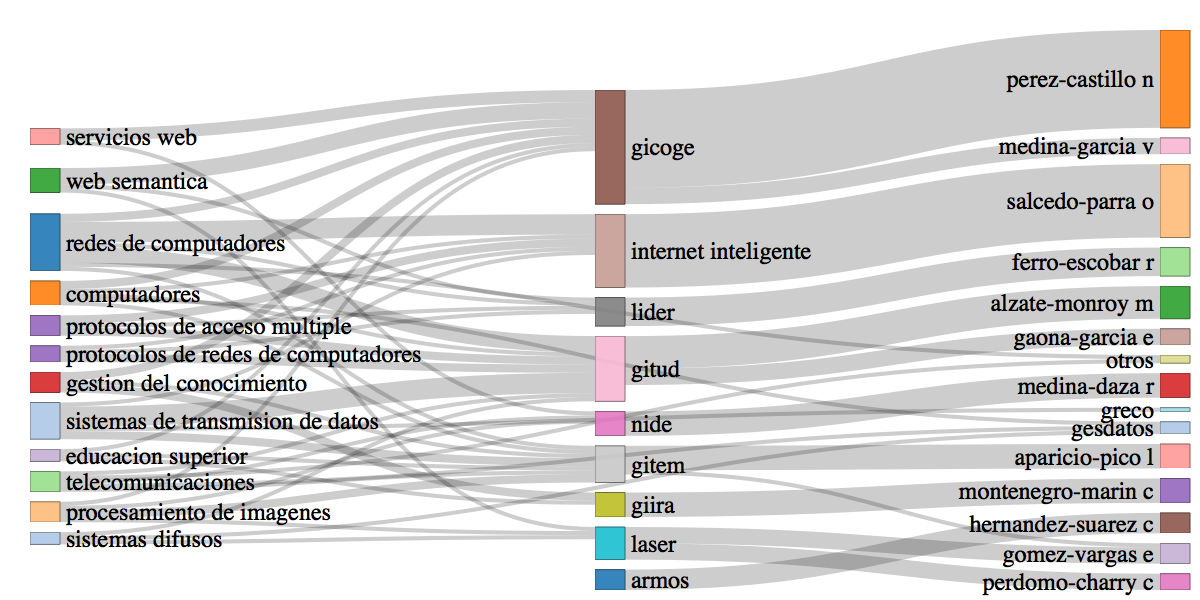}
    \caption{Energy flow through social structures (MIS dataset).}
    \label{cic:energy-flow-social}
\end{figure}


\end{document}